# 3D PHYSICS AND THE ELECTROWEAK PHASE TRANSITION: A FRAMEWORK FOR LATTICE MONTE CARLO ANALYSIS

K. Farakos[a1], K. Kajantie[b2], K. Rummukainen[c3] and M.Shaposhnikov[d4]

[a]*National Technical University of Athens, Physics Department, Zografou Campus, GR 157 80, Athens, Greece*

[b]*Department of Theoretical Physics, P.O.Box 9, 00014 University of Helsinki, Finland*

[c]*Indiana University, Department of Physics, Swain Hall West 117, Bloomington IN 47405 USA*

[d]*Theory Division, CERN, CH-1211 Geneva 23, Switzerland*

## Abstract

We discuss a framework relying on both perturbative and non-perturbative lattice computations which will be able to reliably determine the parameters of the EW phase transition. A motivation for the use of 3d effective theory in the lattice simulations, rather than the complete 4d one, is provided. We introduce and compute on the 2-loop level a number of gauge-invariant order parameters – condensates, which can be measured with high accuracy in MC simulations. The relation between $\overline{\rm MS}$ and lattice condensates is found, together with the relation between lattice couplings and continuum parameters (the constant physics curves). These relations are exact in the continuum limit.

[1]kfarakos@atlas.central.ntua.gr
[2]kajantie@phcu.helsinki.fi
[3]kari@trek.physics.indiana.edu
[4]mshaposh@nxth04.cern.ch

# 1 Introduction

The infrared problem in the thermodynamics of Yang-Mills fields prevents a purely perturbative study of the electroweak phase transition. It is well known that the effective potential for the scalar field cannot be computed perturbatively for small $\phi$, in the symmetric phase. Accordingly, all the quantities characterizing the phase transition region, $T_c$, metastability and tunnelling temperatures, $\phi_c$, latent heat and interface tension are not computable in perturbation theory. At the same time, some properties of the high energy particle excitations in broken and unbroken phases allow perturbative treatment. Moreover, the broken phase of the electroweak theory is weakly coupled, provided the vacuum expectation value of the Higgs field is large enough. Thus it is essential to apply both perturbative and non-perturbative methods to determine the nature of the electroweak (EW) phase transition.

These general remarks suggest that probably the best way to determine the parameters of the electroweak phase transition relevant for cosmology with sufficient accuracy should consist in two steps, separating the perturbative and the non-perturbative physics. At the first step one goes as far as possible with analytical perturbative calculations, which can simplify greatly the underlying 4d EW theory. The second consists in numerical lattice Monte Carlo simulations of non-perturbative physics. This program was initiated in ref. [1] and consists in the following ingredients:

(i) Dimensional reduction of 4d high temperature theory. This step provides an effective 3d theory containing one (or even two) essential mass scales less than the underlying 4d theory. The scales of the 4d theory are the temperature $T$, the Debye screening mass $m_D \sim gT$, and the infrared scale $g_3^2 = g^2 T$, while those of the reduced theory are $m_D$ and $g_3^2$ (or just $g_3^2$ if $A_0$ – the temporal component of the gauge field – is integrated out). This part of the analysis has been carried out in [2] on the 2-loop level; the 2-loop contribution is essential both for numerical and conceptual reasons. The 3d theory is an excellent approximation to the 4d world for realistic Higgs masses. At the same time, this theory is much easier to analyze with perturbative or non-perturbative methods. The main reason is its super-renormalizable character. In contrast to the 4d theory, where ultraviolet divergences exist in any order of perturbation theory, in 3d only 1- and 2-loop graphs are divergent. This makes the scaling behaviour, essential for relating lattice and continuum parameters, much simpler than that in 4d, and allows one to make computations with high accuracy.

(ii) A relation between lattice couplings and couplings of the 3d theory, which is exact in the continuum limit. With this relation, and using (i), one can relate any of the lattice observables to physical ones for given Higgs and $W$ masses. The existence of an exact relation is a consequence of the super-renormalizable character of the 3d gauge-Higgs system, where 3-loop or higher order terms do not contribute to renormalization.

(iii) The study of different observables on the lattice in the broken phase (below the critical temperature) and a comparison of them with 2-loop perturbative predictions [5]. This allows one to estimate higher (3-loop, etc.) order perturbative corrections to-gether with an estimate of finite $a$ (lattice spacing) effects. This part of the framework is quite important, since it can provide a justification for the use of perturbative methods in the broken phase for concrete values of the Higgs mass and the temperature. If perturbation theory works with sufficient accuracy, then in some sense at least one half of the problems associated with the electroweak phase transition – the properties of the broken phase at $T_c$ – can be dealt with analytical methods.

---

[5] As we have shown in [2] the 2-loop level of perturbative computation is a minimal one for which the result is practically scale-independent.

(iv) The study of different observables in the symmetric phase (above $T_c$). Here the most interesting characteristics are correlation lengths of operators with different quantum numbers, in particular those that would be associated with Higgs and $W$ bosons in the broken phase. The results of this study cannot be reproduced by perturbative methods, since the symmetric phase is in the strongly coupled confining phase. This part of the study is important since most of the mechanisms for electroweak baryogenesis are linked in one way or another with the properties of the symmetric phase.

(v) As a combination of (iii) and (iv), the study of the system in the transition region and the determination of the parameters of the phase transition (such as its order, the critical temperature, the jump of the order parameter, etc.).

This paper sets up points (ii) and partially (iii) of the above proposal, and adds several remarks to (i). These steps are necessary for the detailed study of the EW phase transition itself [3]. Some qualitative results on the EW phase transition has appeared in [1] (mostly for the light Higgs boson, $m_H = 35$ GeV) and in [4] for $m_H = 80$ GeV.

The paper is organized as follows. In section 2 we provide further motivation of step (i), i.e. the use of a 3d effective theory [2, 4],[5] rather than the full 4d one [6]–[10]. In section 3 we summarize the results from the 2-loop perturbative analysis [2]. In sections 4 and 5 we introduce a set of gauge-invariant observables (condensates), which can be computed with a simple manipulation from the 2-loop effective potential and can be measured on a lattice with high precision. A discussion of the renormalization of the ground-state energy $\varepsilon_\mathrm{vac}$ in the $\overline{\mathrm{MS}}$ scheme is given in section 4. As a new perturbative result, section 5 contains a 2-loop perturbative computation of $\varepsilon_\mathrm{vac}$ and $\langle \phi^\dagger \phi \rangle$ in the $\overline{\mathrm{MS}}$ scheme.

The lattice action is given in Section 6, as well as a detailed discussion of relating lattice numbers to physical quantities, i.e. defining the constant physics curve. It is shown that the relation of the lattice to continuum requires the computation of three numbers, $\eta$, $\bar\eta$ and $\tilde\eta$. Two of these are computed in 2-loop lattice perturbation theory[6]. The third one, $\eta$, is related to the pure gauge sector, in which lattice perturbation theory is very complicated. We determine it in Section 7 by measuring $\langle \phi^\dagger \phi \rangle$ on the lattice deep in the broken phase and by comparing this with the 2-loop perturbative result. This comparison involves the constant physics curve and permits one to calibrate $\eta$. The constant physics curves we found do not have any higher loop contributions[7]. Simultaneously, we use Monte Carlo data to estimate higher order corrections to the effective potential and the magnitude of finite-$a$ effects. In Section 8 we construct the constant physics curves for a theory in which the $A_0$ field – the former time component of the gauge field – is integrated out. Section 9 is a conclusion. Several technical points are covered in Appendices. In particular, we derive the 2-loop form of the gauge-invariant effective potential introduced in [13].

---

[6] In analogy to the derivation of the relation between $\Lambda_\mathrm{QCD}$ in lattice and continuum regularization schemes [11, 12] in QCD, the calculation amounts to computing the constant term besides the leading logarithmic term. However, now the logarithmic term arises only at the 2-loop level, which accounts for the greater complexity of the problem.

[7] In [1] the corresponding relation was obtained at the 1-loop level.

In this paper we frequently use the results of ref. [2]. All references to specific formulae from that paper are indicated by a I followed by the equation number.

## 2  3d versus 4d

In the following subsection we discuss the basic properties of dimensional reduction. We also introduce a criterion for a dimensional reduction different from that previously discussed in the literature and discuss the role of fermions. In the second subsection we discuss the quantitative advantages of the 3d formulation in the study of the electroweak phase transition.

### 2.1  Why the 3d description is possible

The idea of dimensional reduction [14]–[19] comes from a statement that *equilibrium* finite temperature field theory is equivalent to a Euclidean zero temperature field theory defined on a finite imaginary time interval $1/T$ supplied with periodic boundary conditions for bosons and antiperiodic for fermions. Therefore, 4d finite temperature field theory is exactly equivalent to a 3d field theory at $T = 0$, but with an infinite number of fields. The 3d masses of bosons are $m_B = 2\pi nT$ and those of fermions $m_F = \pi T(2n+1)$, $n = 0, 1, ....$

The 3d effective bosonic action is defined as

$$\exp(-S_{\text{eff}}) = \int D[\text{heavy modes}] \exp(-S), \qquad (1)$$

where the functional integral is taken over all modes with masses $\sim T$. The effective action is a complicated functional of 3d bosonic fields. The use of it is, of course, equivalent to the full 4d theory.

Suppose now that we are dealing with infrared physics only, so that the typical energy scale (which we denote as $Q$ and discuss in more detail later) of the problem is much smaller than the temperature. Then *all* 3d fermionic modes and all bosonic modes besides the one corresponding to $n = 0$ have masses larger than $\pi T$ and are very heavy in comparison with our scale. So, the integration over the heavy modes can be done perturbatively with an expansion parameter $Q/(\pi T)$. The effective 3d action can be represented in the form

$$S_{\text{eff}} = cVT^3 + \int d^3x L_{\text{eff}}(T) + \sum_n \frac{O_n}{T^n}, \qquad (2)$$

where $L_{\text{eff}}(T)$ is a renormalizable 3d effective bosonic Lagrangian with temperature-dependent constants, $O_n$ are operators of dimensionality $n$, suppressed by powers of temperature, $c$ is a number related to the number of degrees of freedom of the theory and $V$ is the volume of the system. For example, the 3d gauge coupling is $g_3^2 = g^2 T$. The last step of dimensional reduction amounts to neglecting the terms $O_n$. Note that the constant part of the action is irrelevant for a discussion of phase transitions.

Formally, the operators $O_n$ are suppressed by the powers of temperature. It seems, therefore, that their contribution is negligible in the limit $T \to \infty$. This is, however, wrong since the 3d coupling and mass also grow without bounds in this limit. Thus $O_n \propto T^n$, and the extra contributions to the 3d action do not vanish when $T \to \infty$. Thus, the Landsman [17] criterion of *exact* dimensional reduction

$$\lim_{T\to\infty} (V_{4d} - T V_{3d})/T^4 \to 0, \qquad (3)$$

where $V_{4d}$ is the effective potential computed in 4d and $V_{3d}$ is the potential computed with a renormalizable part of 3d action (with temperature-dependent parameters), does not hold. Instead, one has

$$(V_{4d} - T V_{3d})/T^4 = \mathcal{O}(m_i^2(T)/T^2), \qquad (4)$$

where the $m_i(T)$ are the relevant mass scales (inverse screening and correlation lengths) of the system. The question of validity of dimensional reduction thus becomes a pragmatic one: when is the r.h.s. of (4) negligible? At very high $T$ all masses are proportional to coupling constants times $T$ and the r.h.s. of (4) is some power of 4d coupling constants. These extra terms are small only when the dimensionless coupling constants (run to the scale of about $7T$, see below) of the underlying 4d theory are small. So, the criterion for dimensional reduction to work is thus the same as that for $T = 0$ perturbation theory: $g^2 \ll 1$ and $\lambda \ll 1$, $\lambda$ being a scalar self-coupling constant. For the electroweak theory (minimal standard model) this implies that $m_H$ should not be very large, say, $m_H \lesssim 3 m_W$.

However, we are mainly interested not in very large $T$ but in $T$ close to a possible phase transition temperature. Then the system contains two phases and may contain other mass scales than those set by the product of $T$ and coupling constants. We will see below that the dimensional reduction is valid for the electroweak theory near $T_c$, provided that the Higgs mass is sufficiently large, say, $m_H \gtrsim 30$ GeV.

One may wonder if we achieved anything going from 4d to 3d: dimensional reduction requires small coupling constants and works only when 4d perturbation theory works. The answer is: Yes, we do. The region of applicability of perturbation theory is different at zero and at high temperatures. The criterion for finite $T$ perturbation theory to work is

$$g_3^2/Q \ll 1 \qquad (5)$$

(for example, in the calculation of the effective potential the condition is $g_3^2/(\pi m_T) \ll 1$, in hot QCD perturbation theory breaks down for distances $1/Q > 1/(g^2 T)$, etc.), while the criterion for $T = 0$ perturbation theory to work, which is the same as the criterion of dimensional reduction to be valid, is $g^2 \ll 1$. In other words, at high temperatures, we can find ourselves in a situation in which 4d finite $T$ perturbation theory does not work ($g_3^2/Q = g^2 T/Q \gg 1$), but dimensional reduction is possible ($g^2 \ll 1$). As we shall see, this is a typical situation in electroweak theory. After the reduction is done analytically, we can do some analytic perturbative or, most importantly, numerical non-perturbative analysis of the 3d theory. This may be easier and more precise than the analysis of the 4d one.

Now, we can assert a different *formal* criterion for the validity of dimensional reduction: take the limit $T \to \infty$, with 3d parameters fixed. Then the difference between 4d and 3d computations always vanishes. This is not a physical limit since it implies a transition between different physical theories at zero temperature ($g^2 = g_3^2/T \to 0$). It is, however, quite helpful from a practical point of view, since it provides a definition of the formal expansion parameter, $Q/T$ with $Q \sim g_3^2$, which must be small for the dimensional reduction be be valid. The important point is that in this limit the resulting 3d theory has a non-trivial dynamics, which can be completely non-perturbative.

A final comment concerns QCD at finite temperatures: one should not expect dimensional reduction to be valid in the vicinity of the QCD phase transition, where the 4d gauge coupling is not small. However, at large enough temperatures, it may be used.

## 2.2  Why a 3d description is more efficient

In the previous subsection we argued that 3d effective theories *can be* used for the non-perturbative study of high temperature phase transitions in weakly coupled field theories. In this subsection we will discuss the qualitative and quantitative advantages of the 3d formulation.

We start from the qualitative features. An important aspect of dimensional reduction concerns the role of fermions. Since the fermion masses $(2n+1)\pi T$ and bosonic non-static mode masses $2\pi nT$ both get large simultaneously, both will give a contribution of the same order. Thus either both can be integrated out in the vicinity of $T_c$, i.e. the 4d→3d reduction is valid, or both have to be included and the 3d effective theory is not accurate[8]. Moreover, a realistic electroweak theory contains chiral fermions, which are impossible at present to put on the lattice. So, they must be integrated out analytically. In other words, 4d bosonic theory, which one can study on the lattice, is in any case an effective theory and not better than a dimensionally reduced 3d one.

The second advantage is that complete 4d perturbative computations of the finite $T$ effective potential are known only in the 1-loop approximation. The existing 2-loop computations [20]–[22] deal with high-temperature asymptotics only, which is equivalent to a 3d computation [2]. So, one is forced to compare the results of 4d lattice simulations with effectively 3d perturbative computations. Why then not carry out 3d simulations?

The last, but not the least, advantage is the great simplification of perturbative computations in the 3d effective theory, in comparison with resummed perturbation theory in 4d. This allowed one to clarify the nature of the large 2-loop logarithmic corrections in the $\overline{\rm MS}$ scheme and to introduce a method of summation of those based on the renormalization group [2].

---

[8]The former is the case for the electroweak theory for an interesting range of $m_H$. The latter holds for QCD, for which the gauge coupling is large at $T_c$. In fact, it is well known that the QCD transition depends sensitively on the fermion content and that, for pure SU(3) gauge theory, the 3d approximation starts to work somewhat above $T_c$, for $T \gtrsim 1.5 T_c$.

We now turn to more quantitative aspects of the use of 3d effective theory.

### 2.2.1 The quality of dimensional reduction in continuum

We demonstrate our point by a study of different approximations to the quantity related to the 1-loop free energy of free bosons at high temperatures, entering the 1-loop evaluation of the effective potential and the computation of the effective scalar mass at high temperatures:

$$I(m,T) = T \sum_{n=-\infty}^{\infty} \int \frac{d^3k}{(2\pi)^3} \frac{1}{k^2 + m^2}, \qquad (6)$$

where $k^2 = \mathbf{k}^2 + (2\pi n T)^2$.

The high temperature expansion of (6) in continuum theory in the $\overline{\rm MS}$ scheme reads:

$$I(m,T) = \frac{T^2}{12} - \frac{mT}{4\pi} - \frac{m^2}{8\pi^2} \log(\frac{\mu_T}{T}) + \frac{\zeta(3)}{128\pi^4}\frac{m^4}{T^2} + \mathcal{O}(\frac{m^6}{T^4}), \qquad (7)$$

where $\mu_T = 4\pi T e^{-\gamma} \approx 7T$. The procedure of dimensional reduction takes into account the first three terms here. Namely, the first term changes the zero-temperature mass to the effective high-temperature mass, the third term fixes the scale at which the 4d coupling constants should be normalized, and the second term is purely 3-dimensional (it comes from the $n = 0$ term in the sum). The $O(m^4)$ term and the higher-order terms are omitted in the effective 3d theory. They represent the contribution of higher dimensional operators to the effective action.

By comparison of the second and the fourth terms in the expansion we see that the 3d description is accurate within, say, a 1% level, provided that

$$\frac{\zeta(3)}{32\pi^3}\frac{m^3}{T^3} < 0.01, \qquad (8)$$

i.e. $m/T < 2.0$. This condition is satisfied for all particles in the broken phase in the vicinity of the electroweak phase transition, if the Higgs mass is not very small, $m_H > 30$ GeV. In fact, $m/T$ is smaller than 0.5 for any realistic Higgs mass $m_H > 60$ GeV. This gives an accuracy better than 0.1%. In other words, the 3d description is a very good approximation to the real 4d high-temperature case, at least from the point of view of one-loop computations. The excellent convergence of the high temperature expansion has been mentioned in many papers on this subject (see, for example [23]–[25]).

Of course, the small coefficients in front of higher powers of $(m/T)$, found in one-loop approximation, do not necessarily reappear in 2 or higher loops. The accuracy of the dimensional reduction on the 2-loop level certainly deserves a careful investigation, which can be done on a basis of 2-loop computations in refs. [20, 21, 2, 22].

### 2.2.2 Finite lattice spacing effects

Next we wish to argue that the 4d lattice simulations of the electroweak phase transition can provide the same accuracy in the description of the EW phase transition as 3d simulations, only at the cost of a considerable increase of lattice volume.

We start from a simple heuristic estimate. Let us take for simplicity a 4d lattice with equal lattice spacing $a$ in temporal and spatial directions. The typical momentum of the particle in the plasma is about $3T$, and to have an adequate description of that on the lattice we must have $a \ll 1/(3T)$. In contrast, in 3d simulations the only requirement is that the lattice spacing should be much smaller than a typical correlation length, $a \ll 1/m$, with $m$ being a particle mass. For example, for the theory with Higgs mass $m_H = 80$ GeV the ratio of $W$ mass (defined as $gv(T)/2$ and measured on the lattice) to the temperature at the transition point is of the order of $m/T \sim 0.2$, so that the 3d requirement is about $3/0.2 = 15$ times weaker than that for 4d simulations. Hence, in order to have the same finite $a$ and finite volume effects, one should use $15^3 \sim 3 \cdot 10^3$ larger lattices for 4d simulations.

This rough estimate can, in fact, be confirmed by an analytic study of finite $a$ effects in the quantity $I(m,T)$. For simplicity we take a spatially infinite lattice. Then, in a dimensionally reduced theory we have

$$I_{\text{dimred}}(m,T) = \frac{T^2}{12} + TI - \frac{m^2}{8\pi^2}\log(\frac{\mu_T}{T}). \tag{9}$$

The function $I$, being a 3d tadpole graph, has the following expansion (see Appendix A):

$$I = \frac{1}{4\pi a}[\Sigma - (am) - \xi(am)^2 + \mathcal{O}(am)^3], \tag{10}$$

with $\Sigma = 3.17591$ and $\xi = 0.15281$ [9]. Apart from counterterms, (9) differs from the continuum expression by two terms. First, it does not contain the $O(m^4/T^2)$ term, dropped by the dimensional reduction procedure. As we argued above, this term is negligibly small for the realistic case of electroweak theory. Second, it contains terms proportional to $a$, which vanish in the continuum limit. We denote the magnitude of the finite $a$ correction by $p$. To have, say, $p = 10\%$ (3%), finite $a$ corrections to the 3d contribution to $I_3$, the lattice should obey the condition $1/(am) > \xi/p \simeq 1.5(5)$, which is quite easy to realize in practice.

The 4d lattice approximation to (6) on an asymmetric lattice, which has a finite number of steps $N_t = (aT)^{-1}$ in the temporal direction, is

$$I(m,T,N_t) = T \sum_{n=0}^{N_t-1} \int_{-\pi/a}^{\pi/a} \frac{d^3k}{(2\pi)^3} \frac{1}{m^2 + (\frac{2}{a})^2 \sum \sin^2(\frac{k_j a}{2}) + (\frac{2}{a})^2 \sin^2(\frac{\pi n}{N_t})}. \tag{11}$$

The large $N_t$ (small $a$) expansion of this quantity reads:

$$I_{\text{4dlatt}}(m,T,N_t) = \frac{1}{4\pi}\left(\frac{1.94695}{a^2}\right) + \frac{T^2}{12}[1 + f_1(N_t)]$$

---

[9]Contrary to the 4d case, where finite $a$ corrections scale like $a^2$, scaling violations start from linear in $a$ terms in 3d.

$$+ T(I_3 - \frac{\Sigma}{4\pi a} + \frac{\xi a m^2}{4\pi}) \quad (12)$$

$$- \frac{m^2}{8\pi^2}[\log(N_t) - 0.057 + f_2(N_t)] + \frac{\zeta(3)}{128\pi^4}\frac{m^4}{T^2}[1 + f_3(N_t)] + O(m^6),$$

where the functions $f_1(N_t), f_2(N_t)$ and $f_3(N_t)$ vanish in the continuum limit $N_t \to \infty$ as $\mathcal{O}(1/N_t^2)$ and can be computed numerically (see Table 1).

| $N_t$ | 2 | 3 | 4 | 5 | 6 | 7 | 8 | 9 | 10 |
|---|---|---|---|---|---|---|---|---|---|
| $f_1$ | 0.200 | 0.108 | 0.056 | 0.032 | 0.021 | 0.015 | 0.011 | 0.009 | 0.007 |
| $f_2$ | 0.336 | 0.144 | 0.076 | 0.047 | 0.031 | 0.023 | 0.018 | 0.014 | 0.009 |

Table 1: The values of functions $f_1$ and $f_2$ for different numbers of sites $N_t$ in the temporal direction.

Let us first take for simplicity a lattice minimal subtraction scheme in which only the terms singular in $a$ are removed by counterterms. The formal advantage of 4d is that the scaling violation is proportional to $a^2 \sim \frac{1}{N_t^2}$, and not to $a$, as in 3d. So, the functions $f_1$ and $f_2$ decrease quite rapidly with increase of $N_t$. However, the numerical smallness of the coefficient $\xi$ (reflecting the qualitative discussion of the different scales above), together with the smallness of the ratio $m/T$ near the phase transition, makes the 3d case more advantageous than the 4d one. Indeed, to have the same level $p$ approximation to 4d physics, one should have at least

$$f_1(N_t) < \frac{12}{4\pi} p \frac{m}{T}. \quad (13)$$

For example, for $p = 10\%$ we have $N_t > 5$, and for $p = 3\%$ $N_t > 8$. Of course, this is not too difficult to achieve with computers available at present. However, at the same time, this requirement puts a rather strong limitation on the size of the lattice in the spatial direction. Indeed, for, say, $p = 10\%$ (3%) and $m/T \simeq 0.3$ we have for the correlation lengths $1/(am) \simeq 16(55)$. In other words, to achieve in 4d the same accuracy of description of the EW phase transition as a 3d dimensionally reduced theory can provide, one should use about $10^3$ larger 3d volumes. In total, counting the number of sites in the 4th direction, the use of 3d theory provides a 3 to 4 orders of magnitude gain in computer time[10].

The way out of this argument is the use of the temperature-dependent subtraction scheme, removing the finite contributions to $I_{4\mathrm{dlatt}}(m, T, N_t)$, proportional to $f_1$ and $f_2$. This requires a careful study of the finite $a$ behaviour of observables in 4d simulations at zero and non-zero temperatures[11]. It is, however, highly non-trivial (but in principle possible [9]), since 4d theory contains at least three different mass scales (temperature, Debye screening mass, and 3d coupling $g^2 T$).

---

[10]If the computer time is not a problem, then the 4d simulations with, say, $N_t = 8$, may be superior to 3d simulations, provided the 2- and higher-loop dimensional reduction corrections come with coefficients so large that these corrections are larger that 4d finite $a$ effects.

[11]We are grateful to I. Montvay for a discussion on this point.

# 3  Survey of continuum results

It this section we summarize the continuum formulae relating 3d and 4d theories, and present a number of 2-loop relations in 3d. The starting point is the 4d Lagrangian of SU(2) + fundamental doublet Higgs theory

$$L = \frac{1}{4}F^a_{\mu\nu}F^a_{\mu\nu} + (D_\mu\phi)^\dagger(D_\mu\phi) - \frac{1}{2}m^2\phi^\dagger\phi + \lambda(\phi^\dagger\phi)^2. \tag{14}$$

We have thus omitted from the full standard model the U(1) subgroup as well as fermions, which – with some reservation for the top quark – are inessential for the thermal phenomena studied. In fact, fermions can be trivially included in the scheme of dimensional reduction. Their effect is to modify expressions relating 3d parameters with 4d ones, but they do not in any way change the qualitative nature of the effective 3d theory.

The leading principle of our analysis is the replacement of the full 4d theory (14) by an effective 3d theory, obtained by integrating over the non-static modes perturbatively. As we discussed in the previous section, this is a very good approximation to the problem under consideration.

Integrating over the non-static modes to 1-loop accuracy in the $\overline{\text{MS}}$ scheme the following effective action is obtained:

$$S_{\text{eff}}[A_i^a(\mathbf{x}), A_0^a(\mathbf{x}), \phi_i(\mathbf{x})] = \int d^3x \Big\{ \frac{1}{4}F^a_{ij}F^a_{ij} + \frac{1}{2}(D_iA_0)^a(D_iA_0)^a + (D_i\phi)^\dagger(D_i\phi) +$$
$$+ \frac{1}{2}m_D^2 A_0^a A_0^a + \frac{1}{4}\lambda_A(A_0^a A_0^a)^2 + m_3^2\phi^\dagger\phi + \lambda_3(\phi^\dagger\phi)^2 + h_3 A_0^a A_0^a \phi^\dagger\phi \Big\}. \tag{15}$$

The key property of the theory (15) is that all the couplings $\lambda_A$, $\lambda_3$, $h_3$ and $g_3^2$ are 3d renormalization group invariants (there is no ultraviolet renormalization of them). These couplings are given in terms of the 4d couplings by

$$g_3^2 = g^2(\mu_T)T, \tag{16}$$

$$\lambda_3 = T\left[\lambda(\mu_T) + \frac{1}{16\pi^2}\frac{3}{8}g^4(\mu_T)\right], \tag{17}$$

$$h_3 = \frac{1}{4}g_3^2\left[1 + \frac{1}{16\pi^2}\left(12\lambda(\mu_T) + \frac{49}{6}g^2(\mu_T) - \frac{1}{3}g^2(\mu_T)\right)\right], \tag{18}$$

$$\lambda_A = \frac{17g^4(\mu_T)T}{48\pi^2}, \tag{19}$$

where the 4d couplings are run to the scale $\mu_T$ by the standard $\beta$ functions.

On the contrary, the 3d mass operators for the Higgs and $A_0$ fields contain linear and logarithmic divergences. To renormalize the theory, one adds to eq.(15) the counterterm action

$$\delta S = -\int d^3x \Big\{\delta m^2 \phi^\dagger\phi + \frac{1}{2}\delta m_D^2 A_0^a A_0^a\Big\}, \tag{20}$$

where

$$\delta m^2 = f_{1m}\Sigma_{\text{lin}} + f_{2m}\Sigma_{\text{log}}, \qquad \delta m_D^2 = f_{1D}\Sigma_{\text{lin}} + f_{2D}\Sigma_{\text{log}}. \tag{21}$$

Here
$$f_{1m} = \frac{9}{4}g_3^2 + 6\lambda_3, \qquad f_{1D} = 5(g_3^2 + \lambda_A), \tag{22}$$

and
$$f_{2m} = \frac{81}{16}g_3^4 + 9\lambda_3 g_3^2 - 12\lambda_3^2, \qquad f_{2D} = 5\lambda_A^2 - 20g_3^2\lambda_A \tag{23}$$

are exact expressions with no higher-order corrections and $\Sigma_{\text{lin}}$ and $\Sigma_{\text{log}}$ are counterterms depending on the regularization scheme used. The linearly divergent counterterm $\Sigma_{\text{lin}}$ vanishes in dimensional regularization; on the lattice it behaves like $\Sigma_{\text{lin}} = \Sigma/4\pi a$ ($\Sigma = 3.1759114$, $a$ is the lattice spacing). The quantity $\Sigma_{\text{log}}$ is related to the logarithmically divergent 2-loop sunset diagram (eq.(22) of [2]). In the $\overline{\text{MS}}$ scheme it is
$$\Sigma_{\text{log}} = \frac{1}{64\pi^2\epsilon} \tag{24}$$

and on the lattice
$$\Sigma_{\text{log}} \propto -\frac{1}{16\pi^2}\log(aT). \tag{25}$$

In the $\overline{\text{MS}}$ scheme the masses are scale-dependent:
$$\mu_3 \frac{\partial m_3^2(\mu_3)}{\partial \mu_3} = -\frac{1}{16\pi^2}f_{2m}, \quad \mu_3 \frac{\partial m_D^2(\mu_3)}{\partial \mu_3} = -\frac{1}{16\pi^2}f_{2D}, \tag{26}$$

which can be integrated to give
$$m_3^2(\mu_3) = \frac{1}{16\pi^2}f_{2m}\log\frac{\Lambda_m}{\mu_3}, \quad m_D^2(\mu_3) = \frac{1}{16\pi^2}f_{2D}\log\frac{\Lambda_D}{\mu_3}, \tag{27}$$

where $\Lambda_m$ and $\Lambda_D$ are integration constants. The constant $\Lambda_m$ can be obtained by comparing the 4d and 3d calculations:
$$m_3^2(\mu_3) = \left[\frac{3}{16}g^2(\mu_T) + \frac{1}{2}\lambda(\mu_T) + \frac{g^2}{16\pi^2}\left(\frac{167}{96}g^2 + \frac{3}{4}\lambda\right)\right]T^2 -$$
$$-\frac{1}{2}m^2(\mu_T) + \frac{1}{16\pi^2}\left[f_{2m}\left(\log\frac{3T}{\mu_3} + c\right)\right] = \tag{28}$$
$$\left[\frac{3}{16}g_3^2T + \frac{1}{2}\lambda_3 T + \frac{g_3^2}{16\pi^2}\left(\frac{149}{96}g_3^2 + \frac{3}{4}\lambda_3\right)\right] - \frac{1}{2}m_H^2 + \frac{1}{16\pi^2}\left[f_{2m}\left(\log\frac{3T}{\mu_3} + c\right)\right],$$

where $m_H^2 \equiv m^2(\mu_T)$ and [26]
$$c = [\log(8\pi/9) - 2\gamma_E + \zeta'(2)/\zeta(2)]/2 = -0.348725. \tag{29}$$

For $m_D$ one can similarly write
$$m_D^2(\mu_3) = \frac{5}{6}g^2(\mu_T)T^2 + \frac{1}{16\pi^2}f_{2D}\log(\frac{3T}{\mu_3}) + \frac{1}{16\pi^2}(c_1 g_3^4 + c_2 g_3^2\lambda_3), \tag{30}$$

but the constants $c_1, c_2$ and, hence, $\Lambda_D$ have so far not been computed.

In practice, one can write

$$g(\mu_T) = \frac{2}{3}, \qquad \lambda_3 = \frac{g_3^2 m_H^2}{8 m_W^2}, \qquad m_W = 80.6 \text{ GeV} \qquad h_3 = \frac{1}{4} g_3^2. \qquad (31)$$

The explicit form of the total 2-loop effective potential is needed frequently in what follows. The tree and 1-loop parts are

$$V_0(\phi) = \frac{1}{2} m_3^2(\mu_3) \phi^2 + \frac{1}{4} \lambda_3 \phi^4, \qquad (32)$$

$$V_1(\phi) = -\frac{1}{12\pi}(6 m_T^3 + 3 m_L^3 + m_1^3 + 3 m_2^3) \qquad (33)$$

and the 2-loop part is (eq.(33) of [2])

$$V_2(\phi, \mu_3) = \frac{1}{16\pi^2} \Bigg\{$$
$$-\frac{3 g_3^4}{16} \phi^2 \Big[ 2\bar{H}(m_1, m_T, m_T) - \frac{1}{2} \bar{H}(m_1, m_T, 0) + \bar{H}(m_1, m_L, m_L)$$
$$+ \frac{m_1^2}{m_T^2} [\bar{H}(m_1, m_T, 0) - \bar{H}(m_1, m_T, m_T)]$$
$$+ \frac{m_1^4}{4 m_T^4} [\bar{H}(m_1, 0, 0) + \bar{H}(m_1, m_T, m_T) - 2\bar{H}(m_1, m_T, 0)]$$
$$- \frac{m_1}{2 m_T} - \frac{m_1^2}{4 m_T^2} \Big] - 3 \lambda_3^2 \phi^2 [\bar{H}(m_1, m_1, m_1) + \bar{H}(m_1, m_2, m_2)]$$
$$+ 2 g_3^2 m_T^2 \Big[ \frac{63}{16} \bar{H}(m_T, m_T, m_T) + \frac{3}{16} \bar{H}(m_T, 0, 0) - \frac{41}{16} \Big]$$
$$- \frac{3}{2} g_3^2 [(m_T^2 - 4 m_L^2) \bar{H}(m_L, m_L, m_T) - 2 m_T m_L - m_L^2]$$
$$+ 4 g_3^2 m_T^2 + \frac{3}{8} g_3^2 (2 m_T + m_L)(m_1 + 3 m_2)$$
$$+ \frac{15}{4} \lambda_A m_L^2 + \frac{3}{4} \lambda_3 (m_1^2 + 2 m_1 m_2 + 5 m_2^2)$$
$$- \frac{3}{8} g_3^2 [(m_T^2 - 2 m_1^2 - 2 m_2^2) \bar{H}(m_1, m_2, m_T) + (m_T^2 - 4 m_2^2) \bar{H}(m_2, m_2, m_T)$$
$$+ \frac{(m_1^2 - m_2^2)^2}{m_T^2} [\bar{H}(m_1, m_2, m_T) - \bar{H}(m_1, m_2, 0)]$$
$$+ (m_1^2 - m_2^2)(m_1 - m_2)/m_T + m_T(m_1 + 3 m_2) - m_1 m_2 - m_2^2] \Bigg\}. \qquad (34)$$

Here the masses are

$$m_T = \frac{1}{2} g_3 \phi, \qquad m_L^2 = m_D^2 + \frac{1}{4} g_3^2 \phi^2,$$
$$m_1^2 = m_3^2(\mu_3) + 3 \lambda_3 \phi^2, \qquad m_2^2 = m_3^2(\mu_3) + \lambda_3 \phi^2. \qquad (35)$$

and the finite part of the sunset function is

$$\bar{H}(m_1, m_2, m_3) = \log \frac{\mu_3}{m_1 + m_2 + m_3} + \frac{1}{2}. \tag{36}$$

The coefficient of $\frac{1}{2}\log(\mu_3)\phi^2$, coming entirely from the $H$-terms, gives the value of $f_{2m}/16\pi^2$.

This potential has been calculated in the Landau gauge. In a general covariant gauge the potential has been computed in [27, 28]. The 4d computation of the high temperature asymptotics of the potential [20]–[22] leads to the same result as the 3d one [2].

The 3d scale $\mu_3$ appearing in (28) and in (34) is arbitrary. We refer here to [2] for a discussion of the renormalization group improved effective potential, which allows one to fix this scale in a natural way.

## 4  Gauge-invariant order parameters – condensates

As was argued in ref. [2] a reliable perturbative computation of any physical observable in 3d theory should be done at least on the 2-loop level. The reason for this is the fact that the logarithmic renormalization of masses in 3d starts at the 2-loop level. The only quantity known at present for a 3d gauge-Higgs system on 2-loop approximation is the effective potential in Landau gauge [2] or in arbitrary gauge [27, 28]. The effective potential, is, however, a gauge-dependent quantity, and cannot be immediately used for the extraction of gauge-invariant information. In this section, we show how simple manipulations with the effective potential can generate a set of gauge-invariant order parameters – condensates. Those condensates can be computed with 2-loop accuracy with little extra work and can be easily measured on the lattice.

By condensates, in general, we mean vacuum averages of any local composite gauge invariant operators,

$$\langle O \rangle = \frac{\int \mathcal{D}\psi \, e^{-S(\psi)} O(\psi)}{\int \mathcal{D}\psi \, e^{-S(\psi)}}. \tag{37}$$

Examples are provided by the scalar condensate $\langle \phi^\dagger \phi \rangle$, the "gluon" condensate $\langle F_{ij}^a F_{ij}^a \rangle$, etc. These composite operators contain the product of fields at the same point. Therefore, their expectation values are divergent (infinities cannot be removed by counterterms renormalizing the masses, couplings and fields of the underlying theory). So, their value is prescription dependent and cannot be fixed unambiguously.

We would like to provide a natural prescription for a number of condensates related to the ground state energy. We define its renormalization prescription in the next subsection.

### 4.1  Renormalization of vacuum energy

The renormalization of our 3d theory requires the introduction of 2-loop mass counterterms for $m_3^2$ and $m_D^2$. After that, the computation of any physical amplitude does not contain ultraviolet divergences. However, the computation of the value of the ground state energy defined by the functional integral

$$e^{-V_3 \varepsilon_{vac}} = \int \mathcal{D}\psi \, e^{-S(\psi)}, \tag{38}$$

with $V_3$ being a spatial volume, does contain ultraviolet divergences. We extend the prescription of the $\overline{\text{MS}}$ scheme (remove poles in $\epsilon$) to the vacuum graphs. Thus, vacuum counter-terms are

$$V_2^{ct} = -\frac{1}{16\pi^2 4\epsilon}[6m_D^2 g_3^2 + 3m_3^2 g_3^2],$$
$$V_4^{ct} = -\frac{1}{(4\pi)^4 4\epsilon}[e_1 g_3^6 + e_2 g_3^4 \lambda_3 + e_3 g_3^2 \lambda_3^2 + e_4 \lambda_3^3], \tag{39}$$

($\lambda_A$-terms are, for brevity, not included). Since in the $\overline{\text{MS}}$ scheme linearly divergent integrals are equal to zero, there are no 1- or 3-loop counterterms. After this renormalization the vacuum energy density $\varepsilon_{vac}$ is finite, but $\mu_3$ dependent. For example, the 2-loop expression for the ground state energy of the unbroken phase found from $V_{\text{MS}}(\phi = 0)$ in (34) is:

$$\begin{aligned} V(0) &= \hbar V_1(0) + \hbar^2 V_2(0) \\ &= -\frac{\hbar}{12\pi}[3m_D^3 + 4m_3^3(\mu_3)] \\ &\quad + \frac{3\hbar^2}{16\pi^2}\Big[(2g_3^2 \log \frac{\mu_3}{2m_D} + \frac{3}{2}g_3^2 + \frac{5}{4}\lambda_A)m_D^2 + g_3^2 m_D m_3 + \\ &\quad + (g_3^2 \log \frac{\mu_3}{2m_3} + \frac{3}{4}g_3^2 + 2\lambda_3)m_3^2\Big]. \end{aligned} \tag{40}$$

Clearly, the $\mu_3$ dependence here is a reflexion of the arbitrariness in the definition of $\varepsilon_{vac}$.

The relation of the ground state energy to the effective potential is obvious. The latter is defined through

$$e^{-V_3 V(\phi) - J\phi} = \int \mathcal{D}\psi \, e^{-S(\psi) - J\psi} = e^{-W(J)}, \tag{41}$$

with $\phi = W'(J)$. The external current $J$ is needed to give the scalar field the required value $\phi$. If the external current vanishes, $V'(\phi) = -J = 0$, the system settles in its physical ground state and by evaluating eq.(41) we obtain the value of the effective potential at its minimum, $V(v(T))$, $V'(v(T)) = 0$. This is precisely the ground state energy $\varepsilon_{\text{vac}}$ of the 3d theory defined above by (38).

## 4.2 Renormalized condensates in $\overline{\text{MS}}$ scheme

Now we are ready to define gauge-invariant condensates. Our 3d theory contains five different parameters, $g_3^2, m_3^2, m_D^2, \lambda_3$ and $\lambda_A$. So, there are five special condensates that can be computed by a simple differentiation of the ground state energy density. Since $\varepsilon_{vac}$ is finite, all its derivatives are finite (but $\mu_3$-dependent). So, we *define* the renormalized condensates as corresponding derivatives of $\varepsilon_{vac}$. They are related to the unrenormalized condensates as follows:

1. Quadratic scalar condensates:

$$\langle\phi^\dagger\phi\rangle_R = \frac{\partial\varepsilon_{vac}}{\partial m_3^2} = \langle\phi^\dagger\phi\rangle - \frac{3}{64\pi^2\epsilon}g_3^2, \tag{42}$$

$$\frac{1}{2}\langle A_0^a A_0^a\rangle_R = \frac{\partial\varepsilon_{vac}}{\partial m_D^2} = \frac{1}{2}\langle A_0^a A_0^a\rangle - \frac{6}{64\pi^2\epsilon}g_3^2.$$

2. Quartic scalar condensates:

$$\langle(\phi^\dagger\phi)^2\rangle_R = \frac{\partial\varepsilon_{vac}}{\partial\lambda_3} = \langle(\phi^\dagger\phi)^2\rangle + \frac{\partial f_{2m}}{\partial\lambda_3}\frac{\langle\phi^\dagger\phi\rangle}{64\pi^2\epsilon} + \frac{\partial V_4^{ct}}{\partial\lambda_3}, \tag{43}$$

$$\frac{1}{4}\langle(A_0^a A_0^a)^2\rangle_R = \frac{\partial\varepsilon_{vac}}{\partial\lambda_A} = \frac{1}{4}\langle(A_0^a A_0^a)^2\rangle + \frac{\partial f_{2D}}{\partial\lambda_D}\frac{\langle A_0^a A_0^a\rangle}{128\pi^2\epsilon} + \frac{\partial V_4^{ct}}{\partial\lambda_A}.$$

3. Gauge condensate:

$$-\frac{1}{4}\langle F_{ij}^a F_{ij}^a\rangle_R = g_3^2\frac{\partial\varepsilon_{vac}}{\partial g_3^2} = -\frac{1}{4}\langle F_{ij}^a F_{ij}^a\rangle + g_3^2\frac{\partial f_{2m}}{\partial g_3^2}\frac{\langle\phi^\dagger\phi\rangle}{64\pi^2\epsilon}$$

$$+g_3^2\frac{\partial f_{2D}}{\partial g_3^2}\frac{\langle A_0^a A_0^a\rangle}{128\pi^2\epsilon} + g_3^2\frac{\partial(V_2^{ct}+V_4^{ct})}{\partial g_3^2}. \tag{44}$$

All renormalized condensates defined above are finite but $\mu_3$-dependent.

For completeness, we also write here the relations between the condensate of the kinetic part of the action with other unrenormalized condensates. These follow from the independence of the ground state energy on the normalization of the $\phi$ and $A_0$ fields in the functional integral (38):

$$\langle(D_i\phi)^\dagger(D_i\phi) + (m_3^2+\delta m^2)\phi^\dagger\phi + 2\lambda_3(\phi^\dagger\phi)^2 + h_3 A_0^a A_0^a \phi^\dagger\phi\rangle = \text{const},$$

$$\langle(D_i A_0^a)^\dagger(D_i A_0^a) + (m_D^2+\delta m^2)A_0^a A_0^a + \lambda_A(A_0^a A_0^a)^2 + 2h_3 A_0^a A_0^a \phi^\dagger\phi\rangle = \text{const}, \tag{45}$$

which are nothing but Schwinger–Dyson equations. The constants here do not depend on the parameters of the theory.

The condensates themselves do not have much physical meaning, just because they (and the ground state energy) are dependent on the normalization point. However, if the system exhibits a first-order phase transition, then in the *differences* between condensates in different phases the $\mu_3$ dependence cancels out. An important physical characteristic of the system, namely the latent heat, is related to the jump of the renormalized (or, what is the same, unrenormalized) Higgs scalar condensate. Another application of condensates is that they can be measured on the lattice with high accuracy. Below we will provide a relation between condensates on the lattice and in the continuum.

### 4.3 The latent heat of the phase transition

The computation of the latent heat of the first-order phase transition is based on three equations.
1. The ground state energy density depends on six dimensionful variables and has the dimensionality GeV$^3$. From here we get

$$[g_3^2\frac{\partial}{\partial g_3^2} + \lambda_3\frac{\partial}{\partial \lambda_3} + \lambda_A\frac{\partial}{\partial \lambda_A} + 2m_3^2\frac{\partial}{\partial m_3^2} + 2m_D^2\frac{\partial}{\partial m_D^2} + \mu_3\frac{\partial}{\partial \mu_3}]\varepsilon_{vac} = 3\varepsilon_{vac}. \qquad (46)$$

2. The difference between the ground state energies $\Delta\varepsilon$ of the broken and unbroken phases is renormalization group invariant ($\mu_3$-independent). Therefore

$$\mu_3\frac{d\Delta\varepsilon_{vac}}{d\mu_3} = [\mu_3\frac{\partial}{\partial \mu_3} - \frac{f_{2m}}{16\pi^2}\frac{\partial}{\partial m_3^2} - \frac{f_{2D}}{16\pi^2}\frac{\partial}{\partial m_D^2}]\Delta\varepsilon_{vac} = 0. \qquad (47)$$

3. The derivative of $\Delta\varepsilon$ with respect to the temperature for $g, \lambda$ and $m_H$ fixed [12] at the transition point (where $\Delta\varepsilon = 0$) is just

$$T^2\frac{d\Delta\varepsilon_{vac}}{dT} = m_H^2 T\Delta\langle\phi^\dagger\phi\rangle. \qquad (48)$$

Remembering that, in 4d notation, $\Delta\varepsilon_{vac} = \Delta F/T = -\Delta p(T)/T$, this is nothing but the latent heat $L = \Delta[Tp'(T)]$ of the transition. Here $F$ is the free energy and $p$ is the pressure of the system. According to (42), the jump of this order parameter does not contain any divergences, and, therefore, does not depend on the regularization scheme. The simplicity of expression (48) could be contrasted with the form of the corresponding expression in 4d theory [6].

Another quantity which may be of interest is the difference between the ground state energies in the close vicinity of the phase transition. It is given by

$$\Delta\varepsilon_{vac} = -\Delta\frac{1}{3}\langle\int\frac{d^3x}{V_3}\left[\frac{1}{4}F_{ij}^a F_{ij}^a - 2(m_3^2 + \delta m^2 + \frac{f_{2m}}{32\pi^2})\phi^\dagger\phi - \lambda_3(\phi^\dagger\phi)^2\right.$$
$$\left. -(m_D^2 + \delta m_D^2 + \frac{f_{2D}}{32\pi^2})A_0^a A_0^a - \frac{1}{4}\lambda_A(A_0^a A_0^a)^2\right]\rangle. \qquad (49)$$

We stress that all condensates appearing in this expression are unrenormalized, so that there is no difficulty in putting it on the lattice.

## 5  Perturbative results for the ground state energy and $\langle\phi^\dagger\phi\rangle$

In this section, we shall compute the ground state energy $\varepsilon_{vac} = V(v(T))$, $V'(v(T)) = 0$ and the scalar condensate $\langle\phi^\dagger\phi\rangle$ in 2-loop perturbation theory. The second can be obtained from the first by the use of eq.(42) and the first can be obtained from the 2-loop effective potential (34).

---
[12] We neglect here the slow logarithmic variation of all coupling constants with the temperature.

In perturbation theory the effective potential is derived as an expansion with respect to the Planck constant $\hbar$, which serves as a loop counting parameter:

$$V(\phi) = \sum_{n=0}^{N} \hbar^n V_n(\phi). \tag{50}$$

Simultaneously, the equation for the determination of the vacuum expectation value $v(T)$ is

$$\sum_{n=0}^{N} \hbar^n \frac{dV_n(\phi)}{d\phi}\bigg|_{\phi=v(T)} = 0. \tag{51}$$

We should now solve $v(T)$ from (51) and insert in (50). Two different regimes should be distinguished.

1. **The "classical" regime.** In this case, spontaneous symmetry breaking occurs on the tree level, and the equation $dV_0(\phi)/d\phi = 0$ has a non-trivial solution, $v^2(T) = -\frac{m_3^2}{\lambda_3}$. If the expansion parameter $\rho = g_3^2/(\pi m_T) = \frac{2}{\pi}\sqrt{g_3^2 \lambda_3/(-m_3^2)}$ is small enough, eq. (51) can be solved perturbatively:

$$v(T) = \sum_{n=0}^{N} \hbar^n v_n(T). \tag{52}$$

Inserting this in (50), the ground state energy becomes a power series in $\hbar$ up to the order $N$:

$$\varepsilon_{vac} = \sum_{n=0}^{N} \hbar^n V_n(\phi)|_{\phi=v(T)}, \tag{53}$$

where all terms containing powers of $\hbar$ higher than $N$ should be dropped. It is assumed that $V_n(\phi)$ are re-expanded in powers of $\hbar$. This approximation works provided that

$$m_3^2 < 0, \quad |m_3^2| \gg g_3^2 \lambda_3. \tag{54}$$

2. **The Coleman-Weinberg regime.** Here spontaneous symmetry breaking occurs due to radiative corrections and the equation $dV_0(\phi)/d\phi = 0$ need not have non-trivial solutions. The conditions (54) are not satisfied. In this case a perturbative solution of eq. (51) makes no sense. So, eq. (51) should be solved exactly. We denote the solution by $v_N(T)$. Again, if the parameter $\rho = g_3^2/(\pi m_T) = 2g_3/(\pi v_N(T))$ is small enough, the quantity

$$\varepsilon_{vac} = \sum_{n=0}^{N} \hbar^n V_n(v_N(T)) \tag{55}$$

is a good approximation to the ground state energy. One usually enters the Coleman-Weinberg regime when the tree scalar mass is small enough and when $\lambda_3 \ll g_3^2$. For $m_3^2 = 0$, expression (55) provides an expansion of the vacuum energy with respect to $\lambda_3/g_3^2$; in each order of $\lambda_3/g_3^2$ the summation of all powers of $\hbar$ is automatically performed. Very close to the phase transition the first method certainly fails and the second method should be used. On the contrary, if condition (54) is satisfied, then a Coleman-Weinberg type of computation of the ground state energy can be performed as well. In 3d, if the effective potential is computed up to the terms of order $\hbar^N$, then the difference between the two computations is of the order of $\hbar^{(N+\frac{1}{2})}$. The fractional power of $\hbar$ is due to the contribution of Goldstone bosons to the effective potential, which produces infrared-dangerous terms in the second type of computation. Since fractional powers of $\hbar$ must be absent, the first method of computation should be used in this regime.

Below we shall derive explicit formulas for the ground state energy in the broken phase in the 2-loop approximation in the classical regime. The Coleman-Weinberg type of computation does not require any additional analytic work: simply, one numerically minimizes the 2-loop potential defined in eq.(34) and then determines the ground state energy at the minimum of the potential. Finally, one can use (42) for the determination of condensates.

## 5.1 The ground state energy to order $\hbar^2$

Let us compute $\varepsilon_{\text{vac}} = V(v(T))$ in the classical regime, when spontaneous symmetry breaking occurs already at tree level,

$$\phi_0^2 = \frac{-m_3^2}{\lambda_3}. \tag{56}$$

It is assumed that $-m_3^2 > 0$, and sufficiently large. At this point the masses are

$$\bar{m}_T^2 = \frac{-m_3^2 g_3^2}{4\lambda_3}, \quad \bar{m}_L^2 = m_D^2 + \bar{m}_T^2, \quad \bar{m}_1^2 = -2m_3^2, \quad \bar{m}_2^2 = 0 \tag{57}$$

and the leading approximation to $V(v)$ is:

$$V_0(v) = -\frac{m_3^4}{4\lambda_3}. \tag{58}$$

The corrections to eqs.(56) and (58) can be obtained from the 2-loop result (I.33) for the effective potential as follows. The definition of $v$ is

$$V'(v) = 0, \quad V(v) = V_0(v) + \hbar V_1(v) + \hbar^2 V_2(v) + \mathcal{O}(\hbar^3), \tag{59}$$

where $v^2$ coincides with $\phi_0^2$ only to leading order. To higher orders (note that $v_{(n)}^2$ is the order $\hbar^n$ contribution to $v^2$, not $(v_n)^2$)

$$v^2 = v_{(0)}^2 + \hbar v_{(1)}^2 + \hbar^2 v_{(2)}^2 + \mathcal{O}(\hbar^3). \tag{60}$$

Inserting this in $V'(v^2) = 0$ (prime means derivative with respect to $\phi^2$) and expanding gives the equations

$$m_3^2 + \lambda_3 v_{(0)}^2 = 0, \tag{61}$$

$$\lambda_3 v_{(1)}^2 + 2V_1'(v_{(0)}^2) = 0, \tag{62}$$

$$\lambda_3 v_{(2)}^2 + 2V_1''(v_{(0)}^2)v_{(1)}^2 + 2V_2'(v_{(0)}^2) = 0. \tag{63}$$

For later use, we tabulate the partial derivatives of $V_1$:

$$\frac{\partial V_1}{\partial m_3^2} = -\frac{\hbar}{8\pi}(m_1 + 3m_2), \tag{64}$$

$$\frac{\partial V_1}{\partial \phi^2} = -\frac{3\hbar}{8\pi}\lambda_3[m_1 + m_2 + \frac{g_3^2}{4\lambda_3}(2m_T + m_L)], \tag{65}$$

$$\frac{\partial^2 V_1}{\partial \phi^2 \partial m_3^2} = -\frac{3\hbar}{16\pi}\lambda_3\left(\frac{1}{m_1} + \frac{1}{m_2}\right), \tag{66}$$

$$\frac{\partial^2 V_1}{\partial \phi^2 \partial \phi^2} = -\frac{3\hbar}{16\pi}\lambda_3^2\left[\frac{3}{m_1} + \frac{1}{m_2} + \left(\frac{g_3^2}{4\lambda_3}\right)^2\left(\frac{2}{m_T} + \frac{1}{m_L}\right)\right]. \tag{67}$$

From these one finds

$$v_{(0)}^2 = \phi_0^2 = \frac{-m_3^2}{\lambda_3}, \tag{68}$$

$$v_{(1)}^2 = -\frac{2}{\lambda_3}V_1'(v_{(0)}^2) = \frac{3}{4\pi}\left[\bar{m}_1 + \frac{g_3^2}{4\lambda_3}(2\bar{m}_T + \bar{m}_L)\right], \tag{69}$$

$$\begin{aligned}v_{(2)}^2 &= \frac{4}{\lambda_3^2}V_1''(v_{(0)}^2)V_1'(v_{(0)}^2) - \frac{2}{\lambda_3}V_2'(v_{(0)}^2) \\ &= \frac{9}{32\pi^2}\lambda_3\left[\frac{3}{\bar{m}_1} + \frac{1}{\bar{m}_2} + \left(\frac{g_3^2}{4\lambda_3}\right)^2\left(\frac{2}{\bar{m}_T} + \frac{1}{\bar{m}_L}\right)\right]\left[\bar{m}_1 + \frac{g_3^2}{4\lambda_3}(2\bar{m}_T + \bar{m}_L)\right] \\ &\quad -\frac{2}{\lambda_3}V_2'(v_{(0)}^2).\end{aligned} \tag{70}$$

We have here explicitly written out the "disconnected" parts that follow from the derivatives of the 1-loop potential. The 2-loop expressions follow automatically from eq.(34), but are much lengthier. Note that the second derivatives of $V_1$ in (66,67) contain terms $\sim 1/m_2$, which diverge at the saddle point. Similar terms appear in the first derivatives of $V_2$. Evaluating the first derivative $dV_2/d\phi^2$ one can show that the $1/m_2$-terms cancel between the 1- and 2-loop parts in eq.(70) before taking the limit $m_2 \to 0$. This cancellation actually takes place only in the Landau gauge [27], but this is not surprising since $v(T)$ is not a physical gauge invariant quantity.

With this expansion of $v^2(T)$ we can write

$$\begin{aligned}V(v(T)) &= V_0(v_{(0)}^2 + \hbar v_{(1)}^2 + \hbar^2 v_{(2)}^2) + \hbar V_1(v_{(0)}^2 + \hbar v_{(1)}^2) + \hbar^2 V_2(v_{(0)}^2) = \\ &= V_0(v_{(0)}^2) + \hbar V_1(v_{(0)}^2) + \hbar^2\{V_2(v_{(0)}^2) - \lambda_3^{-1}[V_1'(v_{(0)}^2)]^2\} \\ &= -\frac{m_3^4(\mu_3)}{4\lambda_3} - \frac{\hbar}{12\pi}(6\bar{m}_T^3 + 3\bar{m}_L^3 + \bar{m}_1^3) \\ &\quad + \hbar^2\left[V_2(\bar{m}_T, \bar{m}_L, \bar{m}_1, m_2 = 0) - \frac{9\lambda_3}{64\pi^2}\left(\bar{m}_1 + \frac{g_3^2}{4\lambda_3}(2\bar{m}_T + \bar{m}_L)\right)^2\right].\end{aligned} \tag{71}$$

In other words, the value of $V(v)$ in the loop expansion is given by the value of the 2-loop potential for the saddle point masses in eq.(57) corrected to order $\hbar^2$ by the last term in eq.(71). This corresponds to a set of 1-particle reducible diagrams, which are neglected in the computation of the effective potential.

We shall later also need the 3- and 4-loop contributions to $V(v(T))$. These are

$$V_{(3)}(v(T)) = V_3 + (V_2' + \frac{1}{2}V_1''v_{(1)}^2)v_{(1)}^2, \tag{72}$$

$$V_{(4)}(v(T)) = V_4 + [V_3' + \frac{1}{2}V_2''v_{(1)}^2 + \frac{1}{6}V_1'''(v_{(1)}^2)^2]v_{(1)}^2 - \frac{1}{4}\lambda_3(v_{(2)}^2)^2, \tag{73}$$

where all the potentials should be evaluated at the saddle point as in (71). These equations can be used for an estimate of the higher-order corrections to the effective potential through the lattice measurements of the scalar condensate.

For completeness we also present 2-loop expressions for the unbroken phase. If $m_3^2 > 0$, there is another symmetric phase saddle point at $\phi_0^2 = 0$ with the associated mass values

$$\tilde{m}_T = 0, \quad \tilde{m}_L = m_D, \quad \tilde{m}_1^2 = \tilde{m}_2^2 = m_3^2 \tag{73}$$

and a vanishing leading minimum value

$$V_0(0) = 0. \tag{74}$$

At the point (73) higher-order loop expansion will lead to infrared divergences, but to this order one similarly obtains $v = \mathcal{O}(\hbar^3) = 0$ ($\langle \phi^\dagger\phi \rangle$ is non-vanishing, see below) and $V(0) = \hbar V_1(0) + \hbar^2 V_2(0)$ as given by eq.(40). Note the explicit $\mu_3$-dependence here. As it should, the difference $V(v(T)) - V(0)$ computed from eqs.(71) and (40) is independent of $\mu_3$ to this order in $\hbar$.

## 5.2 Computation of $\langle \phi^\dagger\phi \rangle$ to 2 loops

The result for $\langle \phi^\dagger\phi \rangle$ to 2 loops can now be directly obtained from the expansion (68-70) and the general formula (42). According to (42) we have, in the broken phase ($\int d^3x/V_3$ is implied in the averages),

$$\langle \phi^\dagger\phi \rangle = \frac{\partial(V_0 + V_1 + V_2)}{\partial m_3^2}$$
$$= \frac{1}{2}[v_{(0)}^2 + \hbar v_{(1)}^2 + \hbar^2 v_{(2)}^2] + \frac{\partial V_1(v_{(0)}^2)}{\partial m_3^2} + \frac{\partial^2 V_1(v_{(0)}^2)}{\partial m_3^2 \partial \phi^2}\hbar v_{(1)}^2 + \frac{\partial V_2(v_{(0)}^2)}{\partial m_3^2}$$
$$= \langle \phi^\dagger\phi \rangle_{(0)} + \hbar\langle \phi^\dagger\phi \rangle_{(1)} + \hbar^2\langle \phi^\dagger\phi \rangle_{(2)}, \tag{75}$$

where

$$\langle \phi^\dagger\phi \rangle_{(0)} = \frac{-m_3^2}{2\lambda_3}, \tag{76}$$

$$\langle \phi^\dagger\phi \rangle_{(1)} = \frac{\partial V_1}{\partial m_3^2} - \frac{\partial V_1}{\lambda_3 \partial \phi^2} = \frac{1}{4\pi}\left[\bar{m}_1 + \frac{3g_3^2}{8\lambda_3}(2\bar{m}_T + \bar{m}_L)\right], \tag{77}$$

$$\langle \phi^\dagger\phi \rangle_{(2)} = \frac{1}{2}v_{(2)}^2 + \frac{\partial^2 V_1(v_{(0)}^2)}{\partial m_3^2 \partial \phi^2}v_{(1)}^2 + \frac{\partial V_2(v_{(0)}^2)}{\partial m_3^2}$$

$$
\begin{aligned}
&= \left(-\frac{1}{\lambda_3}\frac{\partial^2 V_1}{\partial \phi^2 \partial \phi^2} + \frac{\partial^2 V_1}{\partial m_3^2 \partial \phi^2}\right)\left(-\frac{2}{\lambda_3}\frac{\partial V_1}{\partial \phi^2}\right) + \left(\frac{\partial V_2}{\partial m_3^2} - \frac{\partial V_2}{\lambda_3 \partial \phi^2}\right) \\
&= \frac{9\hbar^2}{64\pi^2}\lambda_3\left[\frac{2}{\bar{m}_1} + \left(\frac{g_3^2}{4\lambda_3}\right)^2\left(\frac{2}{\bar{m}_T} + \frac{1}{\bar{m}_L}\right)\right][\bar{m}_1 + \frac{g_3^2}{4\lambda_3}(2\bar{m}_T + \bar{m}_L)] \\
&\quad -2\frac{\partial V_2}{\partial m_1^2} - \frac{g_3^2}{4\lambda_3}\left(\frac{\partial V_2}{\partial m_T^2} + \frac{\partial V_2}{\partial m_L^2}\right) - \frac{1}{\lambda_3}\frac{\partial V_2}{\partial \phi^2}. \quad (78)
\end{aligned}
$$

All derivatives here should be evaluated at the saddle point masses in eq.(57); in practice those of $V_2$ are most conveniently evaluated numerically. The 1-loop expression (77) will be written later in a more general form, appropriate for lattice regularization, by using the tadpole function $I(m)$ introduced in Appendix A. Note that here also poles in $m_2$ are generated by the derivatives $dm_2/dm_3^2 = 1/2m_2$ and $dm_2/d\phi^2 = \lambda_3/2m_2$. Because $m_2^2 = m_3^2 + \lambda_3\phi^2$, these precisely cancel in eq.(78) and leave only the terms in the last of eqs.(78). As shown in [27] this cancellation takes place for any gauge fixing parameter $\xi$, as should for a physical quantity.

In the symmetric phase one similarly obtains from eq.(40) that[13]

$$\langle \phi^\dagger \phi \rangle_{\text{symm}} = -\frac{\hbar m_3}{2\pi} + \frac{3\hbar^2}{16\pi^2}\left(g_3^2 \log \frac{\mu_3}{2m_3} - \frac{1}{4}g_3^2 + 2\lambda_3 + \frac{g_3^2 m_D}{2m_3}\right). \quad (79)$$

In a 3d theory the perturbative computation of $\langle \phi^\dagger \phi \rangle$ contains a linear 1-loop and logarithmic 2-loop divergence and these have been treated above in the continuum $\overline{\text{MS}}$ scheme. Due to their importance for the lattice computation it may be useful to specify their origin once more. Their coefficients can be obtained as follows:

$$
\begin{aligned}
\langle 2\phi^\dagger \phi \rangle &= 2\frac{dV(v(T), m_3^2)}{dm_3^2} + \langle \sum_0^3 \phi_a^2 \rangle_{\text{div}} \\
&= 2\frac{dV(v(T), m_3^2)}{dm_3^2} + 4\int \frac{d^3p}{(2\pi)^3}\frac{1}{p^2 + Ap} = \\
&= 2\frac{dV(v(T), m_3^2)}{dm_3^2} + 4\int \frac{d^3p}{(2\pi)^3}\frac{1}{p^2} + \frac{3g_3^2}{4}\int \frac{d^3p}{(2\pi)^3}\frac{1}{p^3} + ... \quad , \quad (80)
\end{aligned}
$$

where $A = -3g_3^2/16$ is the coefficient of the linear term in the 1-loop Higgs self-energy ($\Pi_\phi(k) = Ak$, [2], eq.(83)). In other words, the linear divergence comes from the simple scalar loop and the logarithmic divergence from the linear term in the scalar self energy caused by the emission and absorption of a gauge particle.

The computation of the vacuum expectation value of the composite gauge invariant operator $\phi^\dagger \phi$ is related to the effective potential for this quantity introduced in [13]. We clarify this relation and compute the corresponding effective potential to 2 loops in Appendix B.

---

[13] We would like to note here that since perturbation theory does not work in the unbroken phase this expression cannot be used for any comparison with the lattice results.

## 5.3 Computation of $\langle A_0^a A_0^a \rangle$ to 2 loops

In contrast to the doublet Higgs field $\phi$, the triplet scalar field $A_0^a$ is dynamically inessential and can be integrated over [2]. Due to the relatively large value of the mass $m_D$, the vev of the triplet scalar field is zero. For pure SU(2) gauge theory this has been numerically studied in [29]. Thus the field $A_0^a$ feels the phase transition only in that the field $\phi$ in $m_L^2 = m_D^2 + g^2\phi^2/4$ develops an expectation value $v(T)$. In analogy with $\phi$ in the symmetric phase (eq.(79)) in 2-loop perturbation theory one can write for the finite part

$$\langle A_0^a A_0^a \rangle = 2\frac{dV(v(T), m_D^2)}{dm_D^2} = -\frac{3}{4\pi}m_L +$$
$$+ \frac{3}{16\pi^2}\left\{g_3^2\left[4\log\frac{\mu_3}{2m_L + m_T} - \frac{2m_D^2 - 2m_T^2 - m_T m_L}{(2m_L + m_T)m_L} + \right.\right.$$
$$\left.\left. + \frac{8m_T + m_1 + 3m_2}{m_L}\right] + \frac{5}{2}\lambda_A\right\}. \tag{81}$$

The linear and logarithmic divergences are

$$\langle A_0^a A_0^a \rangle_{\text{div}} = 3\int \frac{d^3p}{(2\pi)^3}\frac{1}{p^2} + \frac{3g_3^2}{2}\int \frac{d^3p}{(2\pi)^3}\frac{1}{p^3}. \tag{82}$$

# 6 The lattice action and the curves of constant physics

## 6.1 The lattice action

The lattice action corresponding to the continuum Lagrangian in eq.(15) has the following form:

$$S = \beta_G \sum_x \sum_{i<j}(1 - \frac{1}{2}\text{Tr}\, P_{ij}) +$$
$$+ \frac{1}{2}\beta_G \sum_x \sum_i [\text{Tr}\, A_0(\mathbf{x})U_i^{-1}(\mathbf{x})A_0(\mathbf{x}+i)U_i(\mathbf{x}) - \text{Tr}\, A_0^2(\mathbf{x})] +$$
$$+ \sum_x \beta_2^A \frac{1}{2}\text{Tr}\, A_0^2(\mathbf{x}) + \sum_x \beta_4^A(\frac{1}{2}\text{Tr}\, A_0^2(\mathbf{x}))^2 + \tag{83}$$
$$+ \beta_H \sum_x \sum_i [\frac{1}{2}\text{Tr}\, \Phi^\dagger(\mathbf{x})\Phi(\mathbf{x}) - \frac{1}{2}\text{Tr}\, \Phi^\dagger(\mathbf{x})U_i(\mathbf{x})\Phi(\mathbf{x}+i)] +$$
$$+ \sum_x [(1 - 2\beta_R - 3\beta_H)\frac{1}{2}\text{Tr}\, \Phi^\dagger(\mathbf{x})\Phi(\mathbf{x}) + \beta_R(\frac{1}{2}\text{Tr}\, \Phi^\dagger(\mathbf{x})\Phi(\mathbf{x}))^2]$$
$$- \frac{1}{2}\beta_H \sum_x [\frac{1}{2}\text{Tr}\, A_0^2(\mathbf{x})\frac{1}{2}\text{Tr}\, \Phi^\dagger(\mathbf{x})\Phi(\mathbf{x})].$$

Here $U_i(\mathbf{x})$ and $P_{ij}$ are the standard link and plaquette variables, the scalar field $\Phi$ on the lattice is related to the continuum scalar field $\phi$ through

$$\Phi = V R_L, \quad R_L^2 = \frac{2a}{\beta_H}\phi^\dagger\phi = \frac{1}{2}\mathrm{Tr}\,\Phi^\dagger\Phi, \tag{84}$$

where $a$ is the lattice spacing, $V$ is a unitary SU(2) matrix, $R_L$ is the radial mode of the Higgs field, and the lattice matrix field $A_0$ is given in terms of continuum field $A_0^a$ as

$$A_0 = \frac{1}{2}ig_3 a \tau_a A_0^a. \tag{85}$$

The main aim of this section is to find the relation between the parameters $\beta_G$, $\beta_H$, $\beta_R$; $\beta_2^A$ and $\beta_4^A$ of the lattice action and the parameters $g_3^2$, $\lambda_3$, $m_3^2$; $\lambda_A$, $m_D^2$ of the 3d continuum theory. The super-renormalizable character of the 3d theory allows one to write the relation in a form that is exact in the continuum limit $N \to \infty$, $a \to 0$, $N/\beta_G \to \infty$. The 3d continuum parameters were related to the physical 4d ones in the previous section.

The lattice–continuum relation is established in several steps:
1. In subsection 6.2 we formulate the general problem of defining the constant physics curves in the 3d gauge-Higgs system. Then (subsection 6.3) we consider simplifications arising in electroweak theory and show that the problem is reduced to the computation of three pure numbers $(\eta, \bar{\eta}, \tilde{\eta})$ – originating from three different classes of 2-loop diagrams contributing to the Higgs self-energy. In subsections 6.4–6.6 we compute two of those numbers ($\bar{\eta}$ and $\tilde{\eta}$) analytically and formulate a way of MC computation of the third one ($\eta$).
2. In subsection 6.7 we work out an exact relationship between the condensates in the lattice regularization scheme and $\overline{\mathrm{MS}}$ scheme.
4. Finally, in Section 7 we use the 2-loop continuum computation in Section 5 and the relationship between different renormalization schemes in subsection 6.7 to determine the constant $\eta$ by comparison of the lattice data with perturbation theory. This allows one also to estimate the magnitude of 3-loop corrections to the effective potential and the magnitude of 2-loop finite $a$ effects.

## 6.2 Constant physics curves: general formulation of the problem

In the tree approximation all the five lattice coupling constants are given in terms of continuum ones by the following equations, which directly follow from the discretization procedure and the form of the continuum Lagrangian:

$$\beta_G = \frac{4}{g_3^2}\frac{1}{a}, \tag{86}$$

$$\beta_R = \frac{1}{4}\lambda_3 a \beta_H^2 = \frac{\lambda_3}{g_3^2}\frac{\beta_H^2}{\beta_G}, \tag{87}$$

$$\beta_4^A = \frac{\lambda_A}{g_3^2}\beta_G, \tag{88}$$

$$m_3^2|_{\text{tree}} = \frac{2(1 - 2\beta_R - 3\beta_H)}{\beta_H a^2}, \tag{89}$$

$$m_D^2|_{\text{tree}} = -2\frac{\beta_2^A}{\beta_G a^2}. \tag{90}$$

The first three relations (86,87,88) are not spoiled by the renormalization procedure and are exact in the continuum limit, while (89) and (90) are modified when radiative corrections are taken into account. As we have discussed in [2], 1-loop and 2-loop counterterms must be added, so that in the continuum limit $a \to 0$ we have the exact relations

$$\frac{2(1 - 2\beta_R - 3\beta_H)}{\beta_H a^2} = m_3^2(\mu_3) - f_{1m}\frac{\Sigma}{4\pi a} - \frac{f_{2m}}{16\pi^2}\log(\frac{6}{a\mu_3}) - \frac{P_{2m}}{16\pi^2}, \tag{91}$$

$$-2\frac{\beta_2^A}{\beta_G a^2} = m_D^2(\mu_3) - f_{1D}\frac{\Sigma}{4\pi a} - \frac{f_{2D}}{16\pi^2}\log(\frac{6}{a\mu_3}) - \frac{P_{2D}}{16\pi^2}, \tag{92}$$

defining the constant physics curves. Here $\Sigma = 3.17159$ (Appendix A) and the coefficients $f_i$ of the linear 1-loop and logarithmic 2-loop terms were given in the section 3. However, one also needs the 2-loop constant terms (when $a \to 0$) $P_{2m}$ and $P_{2D}$ and their determination is our main task. Note that the above relations are, in fact, $\mu_3$-independent, due to the $\mu_3$-dependence of the mass parameters in the minimal subtraction scheme.

Exactly as $f_{2m}$ and $f_{2D}$, the constants $P_{2m}$ and $P_{2D}$ are, for dimensional reasons, quadratic polynomials constructed from the coupling constants $g_3^2, \lambda_3$ and $\lambda_A$. The coefficients in the polynomial $P_i$ can be determined by explicit computation of the 2-loop mass renormalization in lattice regularization. There are at most six dimensionless numbers ($\eta_i$) determining $P_{2m}$:

$$P_{2m} = \eta_g g_3^4 + \eta_\lambda \lambda_3^2 + \eta_A \lambda_A^2 + \eta_{g\lambda} g_3^2 \lambda_3 + \eta_{gA} g_3^2 \lambda_A + \eta_{A\lambda} \lambda_3 \lambda_A. \tag{93}$$

From the structure of the 2-loop scalar mass renormalization graphs, one can see that $\eta_{A\lambda} = \eta_A = 0$, so that four numbers should be determined. In the $A_0$ case we have, in complete analogy:

$$P_{2D} = \rho_g g_3^4 + \rho_\lambda \lambda_3^2 + \rho_A \lambda_A^2 + \rho_{g\lambda} g_3^2 \lambda_3 + \rho_{gA} g_3^2 \lambda_A + \rho_{A\lambda} \lambda_3 \lambda_A. \tag{94}$$

For the same reasons as in the scalar doublet case we have $\rho_{A\lambda} = \rho_\lambda = 0$. So, the general analysis of this 3d theory requires a knowledge of eight pure numbers which fix the relation between the lattice renormalization and $\overline{\text{MS}}$ schemes. To find them, one should compute the renormalization of the Higgs and $A_0$ masses on the 2-loop level in $\overline{\text{MS}}$ and the lattice regularization schemes, and compare the results. Once they are known, there is a one-to-one correspondence between the lattice couplings and continuum parameters which allows one to relate the results of lattice simulations to the 3d effective theory. Then, to relate 3d and 4d high- temperature physics one should use (28) and (30). The two so far unknown constants in (30) are to be computed by comparison of continuum 3d 2-loop renormalization of the Debye screening mass with corresponding 4d computations. To summarize, we have 10 numbers to be determined to relate 3d lattice simulations to 4d high temperature theory.

## 6.3 Simplifications for EW theory

In the context of the EW transition the problem can be simplified, with a good accuracy, by making use of the fact that the coupling constant $\lambda_A = \frac{17g^2}{48\pi^2}g_3^2$ is numerically very small and that there is no $g_3^4$-term in $f_{2D}$. This allows one to use a formal expansion with respect to the 4d coupling constant $g$, assuming for power counting that $\lambda \sim g^2$ and keeping at most the terms $\sim g^4$ in the previous expressions. In this approximation[14] $f_{2D} = 0$, while the polynomials $P_{2m}$ and $P_{2D}$ are defined by five constants $\eta_g, \eta_{g\lambda}, \eta_\lambda$ and $\rho_g, \rho_{g\lambda}$. In more detail, we shall parametrize

$$P_{2m} = \frac{81}{16}g_3^4\eta + 9\lambda_3 g_3^2\bar{\eta} - 12\lambda_3^2\tilde{\eta} + f_{2m}c, \tag{95}$$

where $c$ is the constant defined in eq.(29). This will cancel when $m_3^2(\mu_3)$ is related to 4d physics.

Another simplification takes place when we notice that the mass of the $A_0$ field is parametrically and numerically larger than that of the vector boson and scalar field near the phase transition in the broken phase, at least for a sufficiently heavy Higgs boson [2]. In addition, the $A_0$ mass in the unbroken phase is parametrically larger than the typical 3d mass scale $g_3^2$. So, in principle the $A_0$ field can be integrated out and a theory without the $A_0$ field with modified coupling constants can be formulated [2]. From eqs.(I.51,I.52,I.53) one can see that the uncertainty of the order of $g^4$ in the Debye mass gives an order $g^6$ uncertainty in the gauge and scalar self-coupling constants and an $O(g^5)$ uncertainty in the Higgs mass. These are higher-order terms according to our convention. Therefore, we take $P_{2D} = 0$ and omit the $O(g^4)$ terms from eq.(30).

Note that the formal suppression of the higher-order terms does not necessarily imply numerical suppression. We consider the systematic uncertainties associated with the ignorance of those terms below.

To summarize, the approximate constant physics curves for the electroweak theory are given by eqs.(91,92) with

$$f_{2D} = P_{2D} = \eta_A = \eta_{gA} = \eta_{A\lambda} = 0 \tag{96}$$

and $P_{2m}$ given by eq.(95).

In spite of all these simplifications we still have to compute three numbers, $\eta, \bar{\eta}$ and $\tilde{\eta}$. In the next subsection and Appendix A we carry out an analytic computation of the parameters $\tilde{\eta}$ and $\bar{\eta}$. The one remaining parameter will be determined by a combination of analytical and Monte Carlo methods.

The previous discussion referred entirely to the 3d effective theory. To have an explicit relation between the temperature and lattice parameters, we introduce into eq.(91) the explicit relations between the parameters of the 4d and 3d theories in

---

[14] Formally, $f_{2D}$ is multiplied by $\log\frac{2}{aT} = \log\frac{g^2\beta_G}{2}$, which is singular in the continuum limit. However, in real lattice simulations with $\beta_G < 40$, the value of this log never exceeds 2, and this term may be neglected. We will estimate the possible influence of this term on the results below.

eqs.(28). Using the parametrization (95) the constant physics curve becomes

$$\frac{m_H^2}{4T^2} = \frac{1}{2}\gamma + \frac{1}{(Ta)^2}\left[3 - \frac{1}{\beta_H} + \frac{2\beta_R}{\beta_H} - Ta\frac{f_{1m}}{8\pi\Sigma}\right.$$
$$\left. + (Ta)^2\frac{f_{2m}}{32\pi^2}\log\frac{Ta}{2} + (Ta)^2\frac{f_{2m} - cP_{2m}}{32\pi^2}\right], \quad (97)$$

where

$$\gamma = \frac{3}{16}g^2 + \frac{1}{2}\lambda + \frac{g^2}{16\pi^2}\left(\frac{149}{96}g^2 + \frac{3}{4}\lambda\right). \quad (98)$$

In a more explicit form this is

$$\frac{m_H^2}{4T^2} = \left(\frac{g^2\beta_G}{4}\right)^2\left[3 - \frac{1}{\beta_H} + \frac{m_H^2}{4m_W^2}\frac{\beta_H}{\beta_G} - \frac{9}{8\pi\beta_G}\left(1 + \frac{m_H^2}{3m_W^2}\right)\Sigma - \right.$$
$$-\frac{1}{2}\left(\frac{9}{4\pi\beta_G}\right)^2\left\{\left(1 + \frac{2m_H^2}{9m_W^2} - \frac{m_H^4}{27m_W^4}\right)\log\frac{g^2\beta_G}{2} + \eta + \frac{2m_H^2}{9m_W^2}\bar{\eta} - \frac{m_H^4}{27m_W^4}\tilde{\eta}\right\}\right]$$
$$+\frac{g^2}{2}\left[\frac{3}{16} + \frac{m_H^2}{16m_W^2} + \frac{g^2}{16\pi^2}\left(\frac{149}{96} + \frac{3m_H^2}{32m_W^2}\right)\right]. \quad (99)$$

It is illuminating to define from here $\beta_H$ for large $\beta_G$. One finds that

$$\beta_H(T) = \beta_H(T = \infty) + \left(\frac{2m_H}{3g^2\beta_G T}\right)^2 + \mathcal{O}(\beta_G^{-3}), \quad (100)$$

where

$$\beta_H(T = \infty) = \frac{1}{3} + \frac{1}{\beta_G}\left[\frac{\Sigma}{8\pi}\left(1 + \frac{m_H^2}{3m_W^2}\right) - \frac{m_H^2}{108m_W^2}\right]$$
$$-\frac{1}{9\beta_G^2}\left[\frac{8}{g^4}\gamma - \frac{1}{3}(A - \rho/12)(A - \rho/6) - B\right] + \mathcal{O}(\beta_G^{-3}), \quad (101)$$

where the abbreviations are: $\rho = m_H^2/m_W^2$, $A = 9\Sigma(1 + \rho/3)/8\pi$ and $B =$ second line of eq.(99) without sign and the factor $1/\beta_G^2$. One sees that the $T$-dependence of $\beta_H$ is very simple and that the interval between $\beta_H = 1/3$ and $\beta_H = \beta_H(T = \infty)$ is unphysical.

The constants $\eta$, $\bar{\eta}$ and $\tilde{\eta}$ can be found by computing the 2-loop Higgs mass renormalization in the lattice scheme. Instead of considering only this, we prefer to work in an equivalent language, namely with the effective potential. This will allow us to get simultaneously a number of results concerning the values of different condensates in the ground state. To make the method clear we shall first consider the simplest case, the computation of $\tilde{\eta}$.

For those who are not interested in details of computations we give the result below:

$$\eta = 2.18(6), \bar{\eta} = 1.01, \tilde{\eta} = 0.44, \quad (102)$$

where only MC statistical errors in determination of $\eta$ are shown.

In [4] some combinations of these numbers, relevant for the study of $m_H = 80$ GeV Higgs boson were presented. At that time the quantity $\bar{\eta}$ was computed by MC methods, rather than analytically. Due to a mistake in the computation of parameter $\tilde{\eta}$, made in [4], the numbers in [4] must be corrected as follows: in eq.(11) of [4] $\eta = 2.0$ instead of 1.35, and in eq.(13) $\bar{\eta} = 0.66$ instead of $-4.70$. This change slightly affects the extraction of the critical temperature from the lattice data, but does not affect qualitative results and conclusions of [4], see [3].

## 6.4 The 2-loop lattice effective potential in scalar theory: computation of $\tilde{\eta}$

The constant $\tilde{\eta}$ multiplies the scalar coupling constant. Therefore, it can be computed in pure scalar theory with SU(2) symmetry. To be more general we take the SO(N) scalar theory with the scalar field in a fundamental representation, considered in Appendix B.1 of ref. [2] (the case we are interested in corresponds to $N = 4$). Since the perturbative computations in scalar field theory on the lattice are not so complicated as in gauge theories, the 2-loop lattice effective potential accounting for finite size and finite $a$ effects can be quite easily computed.

The continuum Lagrangian is (in this subsection, $\lambda_3 \to \lambda, m_3 \to m$)

$$\mathcal{L} = \frac{1}{2}(\partial_i \phi_a)^2 + \frac{1}{2}m^2 \phi_a^2 + \frac{1}{4}\lambda(\phi_a^2)^2, \tag{103}$$

and the 2-loop effective potential for the scalar field $\phi_0$ in the $\overline{\text{MS}}$ scheme is:

$$\begin{aligned}
V_{\overline{\text{MS}}}(\phi_0) &= \frac{1}{2}m^2(\mu)\phi_0^2 + \frac{1}{4}\lambda\phi_0^4 - \\
&\quad -\frac{1}{12\pi}[m_1^3 + (N-1)m_2^3] + \\
&\quad +\frac{\lambda}{64\pi^2}[3m_1^2 + 2(N-1)m_1 m_2 + (N^2 - 1)m_2^2] - \\
&\quad -3\frac{\lambda^2}{16\pi^2}\left(\log\frac{\mu}{3m_1} + \frac{N-1}{3}\log\frac{\mu}{m_1 + 2m_2} + \frac{N+2}{6}\right)\phi_0^2,
\end{aligned} \tag{104}$$

where $m_1^2 = m^2(\mu) + 3\lambda\phi_0^2$, $m_2^2 = m^2(\mu) + \lambda\phi_0^2$ and

$$\mu\frac{\partial m^2(\mu)}{\partial \mu} = -\frac{f_{2m}}{16\pi^2}, \qquad f_{2m} = -2(N+2)\lambda^2. \tag{105}$$

In order to find the relation between the $\overline{\text{MS}}$ and the lattice regularization schemes, we have to perform the corresponding computation on the lattice. We take the lattice Lagrangian in the form

$$\mathcal{L}_L = \frac{1}{2}(\Delta_i \phi_a)^2 + \frac{1}{2}m_B^2 \phi_a^2 + \frac{1}{4}\lambda(\phi_a^2)^2, \tag{106}$$

where $\Delta_i$ is a lattice difference in the $i$-th direction and $m_B$ is the bare mass. Then the effective potential is given by ($\hbar$ is a loop counting parameter)

$$V_L = V_{\text{tree}} + \hbar V_{\text{1-loop}} + \hbar^2 V_{\text{2-loop}} \qquad (107)$$

with

$$V_{\text{1-loop}} = J(m_{1L}) + (N-1)J(m_{2L}), \qquad (108)$$

$$V_{\text{2-loop}} = \frac{\lambda}{4}[3I(m_{1L})^2 + 2(N-1)I(m_{1L})I(m_{2L}) + (N^2-1)I(m_{2L})^2]$$
$$-3\lambda^2[H_L(m_{1L}, m_{1L}, m_{1L}) + \frac{N-1}{3}H_L(m_{1L}, m_{2L}, m_{2L})]\phi_0^2, \qquad (109)$$

where $m_{1L}^2 = m_B^2 + 3\lambda\phi_0^2$, $m_{2L}^2 = m_B^2 + \lambda\phi_0^2$, and the functions $I, J$ and $H_L$ are the lattice analogues of the continuum tadpole, 1-loop energy and sunset diagram functions, defined in Appendix A. For small $am$ these functions have the following expansion in powers of the lattice spacing $a$:

$$I(m) = \frac{1}{4\pi a}[\Sigma - (am) - \xi(am)^2 + O((am)^3)], \qquad (110)$$

$$J(m) = \frac{1}{4\pi a^3}[\text{const} + \frac{1}{2}\Sigma(am)^2 - \frac{1}{3}(am)^3 - \frac{1}{4}\xi(am)^4 + O((am)^5)], \qquad (111)$$

$$H_L(m_{1L}, m_{2L}, m_{3L}) = \frac{1}{16\pi^2}\left(\log\frac{6}{a(m_{1L}+m_{2L}+m_{3L})} + \frac{1}{2} + \zeta\right)$$
$$\equiv H(m_{1L}, m_{2L}, m_{3L}) + \frac{1}{16\pi^2}\left(\log\frac{6}{a\mu} + \zeta\right). \qquad (112)$$

where the numbers $\xi$ and $\zeta$ were determined numerically, $\xi = 0.15281$, $\zeta = 0.09$.

Expanding for small $a$ one obtains

$$V_L = \frac{1}{2}m_B^2\phi_0^2 + \frac{1}{4}\lambda\phi_0^4 - \frac{\hbar a\xi}{16\pi}[m_{1L}^4 + (N-1)m_{2L}^4]$$
$$+ \left[\hbar\frac{\Sigma}{8\pi a} - \hbar^2\frac{\lambda\xi\Sigma(N+2)}{32\pi^2}\right][m_{1L}^2 + (N-1)m_{2L}^2]$$
$$- \frac{\hbar}{12\pi}[m_{1L}^3 + (N-1)m_{2L}^3] - \hbar^2\frac{\lambda\Sigma(N+2)}{32\pi^2 a}[m_{1L} + (N-1)m_{2L}]$$
$$+ \hbar^2\frac{\lambda}{64\pi^2}[3m_{1L}^2 + 2(N-1)m_{1L}m_{2L} + (N^2-1)m_{2L}^2]$$
$$- \frac{3\lambda^2}{16\pi^2}[H_L(m_{1L}, m_{1L}, m_{1L}) + \frac{N-1}{3}H_L(m_{1L}, m_{2L}, m_{2L})]\phi_0^2. \qquad (113)$$

With the choice of the following relation between the lattice and continuum masses

$$m_B^2 = m^2(\mu) - \hbar\lambda(N+2)\frac{\Sigma}{4\pi a} + \hbar^2\frac{\lambda^2(N+2)}{8\pi^2}\left[\log\frac{6}{a\mu} + \zeta\right]. \qquad (114)$$

we get

$$V_L = V_{\overline{\text{MS}}} + \hbar \frac{N\Sigma}{8\pi a} m^2. \qquad (115)$$

Note, in particular, that

$$m_{1L}^3 + \hbar \frac{3\Sigma(N+2)}{8\pi a} m_{1L} = \left[ m_{1L}^2 + \hbar \frac{\lambda\Sigma(N+2)}{4\pi a} \right]^{3/2} + O(\hbar^2) = m_1^3 + O(\hbar^2), \qquad (116)$$

that in 2-loop terms $m_{1L}$ and $m_1$ are equivalent and also that the term $\sim a m_{1L}^4 \to$ constant when $a \to 0$. Of course, $m_B^2$ is $\mu$-independent. In terms of the parametrization in eq.(95) we have

$$P_{2m} = f_{2m}(\tilde{\eta} + c), \quad \tilde{\eta} = \zeta - c = 0.44. \qquad (117)$$

This completes an estimate of $\tilde{\eta}$.

## 6.5 The effective potential in $\overline{\text{MS}}$ and lattice regularization schemes

Since the lattice regularization provides nothing but another subtraction scheme in perturbation theory, and since the only divergences of the 3d theory are those related to mass and vacuum energy renormalization, the exact lattice effective potential can differ from the exact $\overline{\text{MS}}$ continuum one by only two types of terms. The first one, multiplying $\phi^2/2$, is associated with the Higgs mass renormalization, and knowing it would give us the unknown constants in $P_{2m}$. The second is a $\phi$-independent piece connected with the renormalization of the ground state energy. Just by power counting, it comes from diagrams up to 4-loop order. It is irrelevant for the study of the phase transitions but will be related to the different condensates introduced in Section 4. Generalizing the result for the scalar theory derived above, the $a \to 0$ limit of the lattice effective potential is thus

$$\begin{aligned} V_L(\phi) &= V_{\overline{\text{MS}}}(\phi) + f_{1m} \frac{\Sigma}{4\pi a} \frac{1}{2} \phi^2 \\ &+ \frac{1}{16\pi^2} [f_{2m}(\log \frac{6}{a\mu} + c) + \frac{81}{16} g_3^4 \eta + 9\lambda_3 g_3^2 \bar{\eta} - 12\lambda_3^2 \tilde{\eta}] \frac{1}{2} \phi^2 \\ &+ V_L^{\text{vac}}. \end{aligned} \qquad (118)$$

In section 4 we have already specified the renormalization of the vacuum energy or the convention for the value of $V_{\overline{\text{MS}}}(\phi = 0)$. This is a nonphysical divergent quantity and its value could thus be fixed at will. Now when it is fixed, any additional lattice effect is contained in $V_L^{\text{vac}}$.

The general form of $V_L^{\text{vac}}$ to different orders in the loop expansion can be fixed as follows. The 1-loop contribution to it, $V_{1L}^{\text{vac}}$, can be found from the 1-loop lattice effective potential, which as a direct generalization of eq.(108) or of eq.(33) is

$$V_{1\text{-loop}}^L = 6J(m_T) + 3J(m_L) + J(m_1) + 3J(m_2), \qquad (119)$$

where the particle masses are given by eqs.(35). To this order the lattice bare mass $m_B^2$ and $m_3^2$ are equivalent. Including the $m^2/a$-term in the expansion of $J(m)$ one has

$$V_{1L}^{\text{vac}} = \frac{\Sigma}{2\pi a}m_3^2 + \frac{3\Sigma}{8\pi a}m_D^2. \tag{120}$$

The 1-loop term comes just from a bare loop and no coupling constants enter. To 2 loops, first powers of $g_3^2$ and $\lambda_3$ enter and the structure of the 2-loop constant term is

$$V_{2L}^{\text{vac}} = \frac{1}{16\pi^2}[m_D^2(a_1 g_3^2 + a_2 \lambda_3) + m_3^2(a_3 g_3^2 + a_4 \lambda_3)]. \tag{121}$$

From the consideration of possible 2-loop diagrams, $a_2 = 0$. Similarly, from the computation of the previous section we find directly that $a_4 = 0$ (in scalar theory the 2-loop lattice potential for $a \to 0$ vanishes when $\phi = 0$, see (115)). The dependence of $a_1$ and $a_3$ on the renormalization scale can be established from the obvious fact that the lattice effective potential or $V_{\text{MS}}(\phi = 0) + V_2^{\text{vac}}$ must be $\mu_3$-independent. From the explicit expression for $V_{\text{MS}}(\phi = 0)$ in eq.(40) we get in the $a \to 0$ limit

$$a_1 \propto -6\log(a\mu), \qquad a_3 \propto -3\log(a\mu). \tag{122}$$

However, the real task is to compute the constant terms here, which will give $\bar{\eta}$.

The 3- and 4-loop ground state contributions are, analogously,

$$\begin{aligned}
V_3^{\text{vac}} &= \frac{1}{(4\pi)^3 a}[d_1 g_3^4 + d_2 g_3^2 \lambda_3 + d_3 \lambda_3^2] \\
V_4^{\text{vac}} &= -\frac{1}{(4\pi)^4}\log(a\mu)[e_1 g_3^6 + e_2 g_3^4 \lambda_3 + e_3 g_3^2 \lambda_3^2 + e_3 \lambda_3^3],
\end{aligned} \tag{123}$$

where $d_i$ and $e_i$ are pure numbers (we omitted for simplicity the contributions proportional to $\lambda_A$). Since there is (for $a \to 0$) nothing to compensate for the dimensionalities of higher powers of coupling constants, there are no higher-loop contributions to $V_{\text{vac}}$.

## 6.6 The constants $a_3, a_1$ and $\bar{\eta}$

The constants $a_3$ and $a_1$ can be found from the computation of the vacuum energy 2-loop diagrams (Fig. 1) in a theory with unbroken symmetry. This is equal to $V_{vac}$. The constant $\bar{\eta}$ can be found from the 2-loop mass operator of the Higgs field containing the product of scalar and gauge coupling constants. In the Landau gauge the corresponding graphs are shown in Fig. 2. We denote this contribution to the mass operator as $\Sigma_{\lambda g}(p)$, where $p$ is the momentum. It is easy to see that

$$\Sigma_{\lambda g}(0) = 3\lambda_3 \frac{\partial V_{vac}}{\partial m_B^2}. \tag{124}$$

From here we get a relation between $a_3$ and $\bar{\eta}$,

$$a_3 = 3[\log(\frac{6}{a\mu}) + \bar{\eta} + c]. \tag{125}$$

The constant $a_1$ is proportional to $a_3$ with a simple symmetry coefficient, i.e. $a_1 = 2a_3$. So, all three constants are related to one-another. We choose to compute the gauge invariant quantity $V_{\text{vac}}$. This can be done in the unbroken phase, since this quantity is by definition $\phi$-independent. The lattice gauge-Higgs vertices are shown in Fig. 3.

The contributions of the two Feynman diagrams are

$$I_{SV} = \frac{3}{2}g_3^2 \int dp\, dq \frac{1}{\hat{q}^2(\hat{p}^2 + m^2)}\left[\sum_i \cos(ap_i) - (1-\xi)\frac{1}{\hat{q}^2}\sum_i \hat{q}_i^2 \cos(ap_i)\right] \qquad (126)$$

and

$$I_{SSV} = -\frac{3}{4}g_3^2 \int dp\, dq \frac{1}{(\hat{p}^2 + m^2)(\hat{k}^2 + m^2)\hat{q}^2}\sum_{ij}\frac{4}{a^2}\sin\frac{a}{2}(2p_i + q_i)\sin\frac{a}{2}(2p_j + q_j)$$
$$\left[\delta_{ij} - \frac{1-\xi}{\hat{q}^2}\frac{4}{a^2}\sin\frac{a}{2}q_i \sin\frac{a}{2}q_j\right], \qquad (127)$$

where

$$\int dp = \int_{-\pi/a}^{\pi/a} \frac{d^3p}{(2\pi)^3}, \qquad \hat{q}_i = \frac{2}{a}\sin\frac{a}{2}q_i, \qquad \hat{q}^2 = \sum_i \hat{q}_i^2, \qquad \hat{k}_i = \hat{p}_i + \hat{q}_i \qquad (128)$$

and $\xi$ is the gauge-fixing parameter. The way to handle the trigonometric factors in the numerator is to separate from them terms that also appear in the denominator. Additional lattice corrections then also appear. Thus, for example,

$$\sum_i \cos ap_i = 3 - \frac{1}{2}a^2(\hat{p}^2 + m^2) + \frac{1}{2}a^2 m^2, \qquad (129)$$

$$\sum_i \hat{q}_i^2 \cos ap_i = \hat{q}^2 - \frac{1}{2}a^2 \sum_i \hat{q}_i^2 \hat{p}_i^2, \qquad (130)$$

and, putting for brevity $a = 2$,

$$\sum_i \sin^2(2p_i + q_i) = \sum_i [2\sin^2 p_i + 2\sin^2(p_i + q_i) - \sin^2 q_i - 4\sin^2 p_i \sin^2(p_i + q_i)]$$
$$= 2(\hat{p}^2 + m^2) + 2(\hat{k}^2 + m^2) - \hat{q}^2 - 4m^2 - 4\sum_i \sin^2 p_i \sin^2(p_i + q_i), \qquad (131)$$

$$\sum_{ij} \sin(2p_i + q_i)\sin(2p_j + q_j)\sin q_i \sin q_j$$
$$= \sum_{ij}[\sin^2(p_i + q_i) - \sin^2 p_i][\sin^2(p_j + q_j) - \sin^2 p_j], \qquad (132)$$

$$\sum_i [\sin^2(p_i + q_i) - \sin^2 p_i]$$
$$= \sum_i [\sin^2 q_i + 2\sin p_i \sin q_i (1 - 2\sin^2(\frac{1}{2}p_i + \frac{1}{2}q_i))]. \qquad (133)$$

With these formulas and symmetries of the integrand one can firstly show that the term multiplying $\xi$ in $I_{SV} + I_{SSV}$ vanishes: this checks the gauge independence of the result. In terms of the lattice integral $I(m)$ one then obtains

$$I_{SV} = \frac{3}{2}g_3^2 I(0)[3I(m) - \frac{1}{2a} + \frac{1}{2}(am)^2 I(m)]. \tag{134}$$

In the $\overline{\text{MS}}$ scheme this vanishes since then $I(0) = 0$. For the SSV diagram one has

$$I_{SSV} = -\frac{3}{4}g_3^2 \Big[ 4I(0)I(m) - I^2(m) - 4m^2 H_L(m,m,0) \\ -a^2 \int dp\, dq \frac{\sum_i \hat{p}_i^2 \hat{k}_i^2}{(\hat{p}^2 + m^2)(\hat{k}^2 + m^2)\hat{q}^2} \Big], \tag{135}$$

where $H_L$ is the lattice sunset function. The problem here is the last term; we find numerically, by expanding for small $m$, that it contains the term $-4m^2 \cdot \delta/16\pi^2$, $\delta = 1.94$ (see Appendix A.2). Thus the total contribution to $V_{\text{vac}}$ is

$$I_{SV} + I_{SSV} = 3g_3^2 [H_L(m,m,0)m^2 + \frac{1}{2}I(0)I(m) + \frac{1}{4}I^2(m) - \delta m^2 + \frac{1}{4}(am)^2 I(0)I(m)]. \tag{136}$$

In the $\overline{\text{MS}}$ scheme this reproduces the $g_3^2 m_3^2$ term in eq.(40). Now the additional terms lead to the final result

$$a_3 = 3[\log(\frac{6}{a\mu}) + \zeta + \frac{1}{4}\Sigma^2 - \delta]. \tag{137}$$

This gives $\bar{\eta} = 1.01$.

The most difficult task is the determination of the constant $\eta$. Its computation with lattice perturbation theory needs quite complicated 3- and 4- gluon vertices. We do not attempt to make the corresponding computation. Instead, we determine this constant with sufficient accuracy in MC simulations. The idea of that computation is simple. One should pick up some quantity which can be perturbatively computed (at least to 2-loop accuracy) and compare it with a result of Monte Carlo simulations. We choose the Higgs scalar condensate, defined below on the lattice, for these purposes. Its 2-loop continuum computation follows directly from the effective potential, while it can be determined by MC simulations with sufficiently high accuracy.

### 6.7 The quadratic condensates on the lattice

Any condensate defined in the $\overline{\text{MS}}$ scheme has its analogue on the lattice. The exact relationship between the lattice and $\overline{\text{MS}}$ condensates follows from the results of this and previous sections. We will give an explicit form for the quadratic scalar condensate only, the relation between higher condensates requires computation of the ultraviolet divergent vacuum graphs in the lattice regularization scheme on the 3- and 4-loop levels (i.e. a computation of the coefficients $d_i$ and $e_i$ in eq.(123)). Using the exact relation between the lattice and continuum effective potential we get for the scalar condensate:

$$\langle \phi^\dagger \phi \rangle_{\text{latt}} = \frac{\Sigma}{2\pi a} + \frac{3}{16\pi^2}[g_3^2(\log\frac{6}{a\mu} + \bar{\eta} + c)] + \langle \phi^\dagger \phi \rangle_R, \tag{138}$$

or, in terms of the radial mode of the lattice scalar field:

$$\beta_H \langle R_L^2 \rangle = \frac{8}{g_3^2 \beta_G} \langle \phi^\dagger \phi \rangle_R + \frac{\Sigma}{\pi} + \frac{3}{2\pi^2 \beta_G} \left( \log \frac{3g_3^2 \beta_G}{2\mu_3} + \bar{\eta} + c \right). \tag{139}$$

Knowing the effective potential, one also can obtain a width of distribution of the order parameter $R^2$ on finite lattices. This point is discussed in Appendix C.

For the $A_0$ condensate we have, in complete analogy:

$$\langle A_0^a A_0^a \rangle_{\text{latt}} = \frac{3\Sigma}{4\pi a} + \frac{3}{4\pi^2} [g_3^2 (\log \frac{6}{a\mu} + \bar{\eta} + c)] + \langle A_0^a A_0^a \rangle_R, \tag{140}$$

or, in terms of the lattice notations

$$\langle \frac{1}{2} \text{Tr} A_0^2 \rangle = \frac{8}{\beta_G^2 g_3^2} \frac{1}{2} \langle A_0^a A_0^a \rangle_R + \frac{3}{\beta_G} \frac{\Sigma}{4\pi} + \frac{3}{\pi^2 \beta_G^2} \left( \log \frac{3g_3^2 \beta_G}{2\mu_3} + \bar{\eta} + c \right). \tag{141}$$

Of course, the lattice condensates are $\mu_3$-independent, but divergent in the continuum limit $a \to \infty$.

We have now defined the relation between $\langle \phi^\dagger \phi \rangle$ in the continuum and on the lattice and computed it to 2 loops in the continuum. The comparison of the continuum computation with MC simulations will allow us to determine completely the constant physics curve. In fact, this comparison will provide information on the magnitude of higher order effects (3 loops, etc.) as well as on the magnitude of 2-loop finite scaling corrections (see below).

## 7 Monte Carlo determination of the constant physics curves

Now we are ready for a Monte Carlo determination of the parameter $\eta$ – the only one left in the relation of the lattice regularization scheme to the continuum one.

Consider the theoretical prediction (139) for $\langle R_L^2 \rangle$. This relation is exact in the continuum limit. In real MC simulations, one has a finite volume together with a finite lattice spacing $a$. So, there are finite size as well as finite $a$ corrections to (139). Finite size corrections are easy to deal with, because in the broken phase with a non-zero mass gap $m$ finite size corrections die exponentially with lattice size, $\sim \exp(-mN)$. Hence, for any fixed $\beta_G$ one can choose a volume large enough so that the value of $\langle R_L^2 \rangle$ is volume independent. In other words, MC simulations can always be done in such a way that finite size corrections are not essential, and for $\beta_G < 40$ the necessary volume is not very large. For the discussion of the probability distribution for the quantity $\langle \phi^\dagger \phi \rangle$ in the finite volume, see Appendix C.

The finite $a$ corrections are power-like and, therefore, more important. A way to improve the situation is to use the analytical *lattice* 1-loop expression for the average $\langle \phi^\dagger \phi \rangle$, generalizing eq.(77):

$$\langle \phi^\dagger \phi \rangle^L_{(1)} = \frac{\Sigma}{4\pi a} - I(N, m_1) + \frac{3g_3^2}{8\lambda_3} \left[ \frac{3\Sigma a}{4\pi} - 2I(N, m_T) - I(N, m_L) \right]. \tag{142}$$

Then the unknown finite $a$ effects appear on the 2-loop level only. With the use of eq.(142) the theoretical prediction for $\langle\phi^\dagger\phi\rangle$ is

$$\langle\phi^\dagger\phi\rangle = -\frac{m_3^2}{2\lambda_3} + \langle\phi^\dagger\phi\rangle_{(1)}^L + \langle\phi^\dagger\phi\rangle_{(2)} + \delta\langle\phi^\dagger\phi\rangle, \qquad (143)$$

where $\delta\langle\phi^\dagger\phi\rangle$ represents the value of the unknown higher order continuum contributions.

Let us analyse in more detail the behaviour of the function $\Delta \equiv \beta_G\beta_H\Delta R^2 = 8/g_3^2 * \Delta\langle\phi^\dagger\phi\rangle$, where $\Delta R^2$ is the difference between the MC and the analytical results (143) (we take $\beta_H$ corresponding to a "classical" regime, where spontaneous symmetry breaking appears on the tree level). There are three mass scales in our theory in the broken phase. Two of them ($m_T$ and $m_1$) are related to each other through $\bar{m}_T/\bar{m}_1 = m_W/m_H$. The third one is the Debye screening mass $m_D$. The general structure of $\Delta$ is:

$$\Delta = f(am_T, am_D) + (g_3^2 a)\psi(\frac{g_3^2}{\pi m_T}; am_T, am_D), \qquad (144)$$

where $f(am_T, am_D)$ represents the 2-loop finite $a$ effects and $\psi$ the 3- and higher-loop finite $a$ effects. Here $\rho = g_3^2/(\pi m_T)$ is the expansion parameter of the 3d theory. The finite scaling behaviour of the function $f$ is quite complicated due to the presence of two different mass scales, $m_T$ and $m_D$, which have different dependence on the scalar mass $m_3$. We notice, however, that if the vev of the Higgs field is large enough, then $m_T \simeq m_D$, and the function $f$ essentially depends on one variable only. In the following we used the constraint $v(T)/T > 3.6$, which ensures that $m_T \simeq m_D$ with an accuracy of $\sim 10\%$. In fact, this condition, being a bit arbitrary, has a weak influence on the parameter $\eta$ extracted from MC simulations. A further simplification comes about when we notice that for these values of $v/T$ the loop expansion parameter is quite small, $\rho \sim 0.1$, and we neglect the 3- and higher-loop finite $a$ effects.

In general, there are three possibilities:
(i) The lattice spacing is so small that finite $a$ effects are not essential. Then the difference between the lattice result and the 2-loop expression is related to the amplitude of the higher-order corrections to the continuum scalar condensate,

$$\Delta_2 = -\frac{8}{g_3^2}\delta\langle\phi^\dagger\phi\rangle, \qquad (145)$$

so that the function $\Delta_2$ depends on the variable $g_3^2/m_T = 2gT/\phi$ only. It must go to zero for $gT/\phi \to 0$. The value of $\eta$ can be found from this requirement. This type of behaviour is expected for a sufficiently small parameter $am_T$, measuring the magnitude of finite $a$ effects.
(ii) Higher-order corrections are so small that their effects on the scalar condensate are negligible. Then the function $\Delta_2$ depends on the variable $am_T$ only and must go to 0 when $am_T/2 = \phi/(gT\beta_G) \to 0$. The value of $\eta$ follows from this requirement. One expects to enter this regime when the expansion parameter $\rho = g_3^2/(\pi m_T)$ is small enough.

(iii) In the most general case one can neglect neither higher- order nor finite $a$ corrections. Then the function $\Delta_2$ depends essentially on two variables, $am_T$ and $g_3^2/m_T$.

We shall study the most general case. So, we need some expression for the finite $a$ effects described by $f(am_T)$ and an ansatz for higher-order corrections to the scalar condensate. The function $f(x)$ can be expanded in a power series in $x$, starting with a linear term:

$$f(x) = Ax + Bx^2 + \ldots. \tag{146}$$

The linear term appears in the computation of the sunset diagram and the figure-of-eight diagram on the lattice (see Appendix A). Moreover, from a computation of the figure-of-eight graph we know that the $(am)^2$ corrections come with a large coefficient, of the order of the coefficient in front of $am$. Therefore, in the analysis of the 2-loop finite $a$ effects it may not be sufficient to consider the linear term only. So, we will keep two terms in the expansion of $f(x)$.

Simultaneously, we will use the following expression for the function $\delta \langle \phi^\dagger \phi \rangle$ accounting for the 3- and 4- loop contributions:

$$\delta \langle \phi^\dagger \phi \rangle = -\frac{g_3^2}{8\lambda_3} \frac{g_3^4}{(4\pi)^3 m_T} (\tilde{\beta} + \tilde{\gamma} \frac{g_3^2}{4\pi m_T}), \tag{147}$$

where $\tilde{\beta}$ and $\tilde{\gamma}$ are some unknown parameters.

To summarize, the "theoretical" prediction for $\langle R_L^2 \rangle$ contains five numbers ($\eta, A, B,$ and $\tilde{\beta}$ and $\tilde{\gamma}$) which we want to determine by comparing the prediction with lattice MC data.

The parameters of the lattice action (83) are fixed as follows. We choose $\beta_G = 12, 20, 24, 32, 40$, thus fixing $a$ (since $g = 2/3$) and $m_H = 80$ GeV fixing $\lambda$. The $A_0$ couplings are fixed using eqs.(88,90) and eq.(92) with $f_{2D} = P_{2D} = 0$. The system then is simulated for various values of $\beta_H$ ($\beta_R$ is fixed by eq.(87)) for lattice sizes $N = 12, 16, 24, 32, 40, 48$. An example of distributions in $R_L^2$ for $\beta_G = 12$, $m_H = 80$ GeV and $N = 24$ is shown in Fig. 4. They show a single peak deep in the broken phase, which develops into a 2-peak structure at the critical value $\beta_{Hc}$. For each peak the value of $\langle R_L^2 \rangle$ with error is computed.

In Fig. 5 we present $\Delta$ as a function of $\phi/(gT\beta_G) = am_T/2$ for the best fit set of parameters for the Higgs mass $m_H = 80$ GeV. The effect of variation of the different parameters of the fit is roughly as follows: the change of $\tilde{\beta}$ and $\tilde{\gamma}$ changes the deviation of the points corresponding to different $\beta_G$ from the single curve, parameters $A$ and $B$ fix the form of that curve, while the change in $\eta$ moves the curve up or down. The fitted parameters are:

$$\eta = 2.18(6), \quad A = 5.09(25), \quad B = -3.40(23), \quad \tilde{\beta} = -162(22), \tag{148}$$

the parameter $\tilde{\gamma}$ characterizing the 4-loop contribution to the effective potential cannot be determined to any good accuracy by this fit; the numbers given above are stable with respect to variation of $\tilde{\gamma}$ within the limits $-700 < \tilde{\gamma} < 700$. The quality of the fit is quite good, $\chi^2 = 38.9/40$. In the computation of the continuum scalar condensate we took $\mu_3 = 1.6 m_T$, the value for which 2-loop corrections to the effective potential are minimized in the vicinity of the minimum.

The coefficient $\eta$ must be independent of the Higgs mass. To check this, we performed analogous simulations for a different Higgs mass, $m_H = 160$ GeV, with the following result ($\beta_G = 20, 32$):

$$\eta = 2.44(29), \quad A = 0.86(30), \quad B = 0.67(19), \quad \tilde{\beta} = -317(84), \tag{149}$$

with $\chi^2 = 11/9$. Again, the parameter $\tilde{\gamma}$ remains undetermined. One can see that the values of all the parameters besides $\eta$ changed a lot, while $\eta$ remains constant within error bars. The value of $\mu_3$, minimizing the 2-loop contribution for this value of the Higgs mass, is $\mu_3 = 0.39 m_T$.

To check the stability of the result with respect to the constraint $v/T > 3.6$ we made a corresponding fit including the simulations with $v/T \simeq 1$. As expected, the quality of the fit is worse than previously[15] (the function $f$ for small $v/T$ depends essentially on two variables rather than one), but the fit parameters are consistent with the numbers given above, i.e.

$$\eta = 2.17(4), \quad A = 5.10(18), \quad B = -3.43(18), \quad \tilde{\beta} = -151(12), \quad \tilde{\gamma} = -98(83). \tag{150}$$

In this case the parameter $\tilde{\gamma}$ is also determined.

It is instructive to compare the finite scaling contributions to $\Delta$, which are determined by this fit through the coefficients $A$ and $B$ with a typical 2-loop contribution $\sim \frac{f_{2m}}{16\pi^2} \frac{4}{\lambda_3 g_3^2} \sim 1$ for $m_H = 80$ GeV. These corrections are of the order of the 2-loop continuum contribution already at $am_T \sim 0.5$.

The knowledge of the parameter $\tilde{\beta}$ allows one to get an idea of the magnitude of the 3-loop corrections to the effective potential, at least at sufficiently large $\phi$, so that $m_T \approx m_D$. We parametrize the 3-loop corrections in this region as

$$V_3 = \frac{\beta}{(4\pi)^3} g_3^4 m_T + V_3^{\text{sing}}, \tag{151}$$

where $V_3^{\text{sing}}$ is the piece of the 3-loop potential, which is singular in the limit $m_2 \to 0$. It can be found from the equations of subsection 6.1, taking into account the condition that the $O(\hbar^3)$ contribution to the ground state is finite. We get:

$$V_3^{\text{sing}} = -\frac{27}{128 m_2} \frac{g_3^4}{(4\pi)^3} (2m_T + m_L + \frac{4\lambda_3}{g_3^2} m_1)^2. \tag{152}$$

Now, to relate $\beta$ and $\tilde{\beta}$ we use the relation of the 3-loop contribution to the scalar condensate $\langle \phi^\dagger \phi \rangle$ through the 3-loop effective potential given in eq.(72). The difference between $\beta$ and $\tilde{\beta}$ appears from the existence of simply connected diagrams contributing to the condensate. We present here only the numerical result of this computation:

$$\tilde{\beta} \simeq \beta - 113. \tag{153}$$

---
[15] The $\chi^2$ for this fit is 67/54 d.o.f, which gives a confidence level of 0.11.

This gives for the 3-loop contribution $\beta = -49(22)$. This number is quite reasonable and agrees with the expectation that the true loop expansion parameter is about $\frac{g_3^2}{\pi m_T}$ [2].

We conclude this subsection by an estimate of the systematic uncertainty in the determination of $\eta$ due to the poor knowledge of the 2-loop corrections to the Debye screening mass[16]. From eq.(I.53) the change of the effective scalar mass $m_3^2$ due to the change of the Debye mass $m_D$ and parameter $\eta$ is

$$\delta m_3^2 = -\frac{g_3^4}{16\pi^2}\frac{81}{16}\delta\eta - \frac{3g_3^2 \delta m_D^2}{32\pi m_D}. \tag{154}$$

Using eq.(92) the variation of the Debye mass is

$$\delta m_D^2 = -\frac{g_3^4}{16\pi^2}(\rho_g + \frac{\lambda_3}{g_3^2}\rho_{g\lambda}). \tag{155}$$

Therefore,

$$\delta\eta = \frac{1}{54\pi}\frac{g_3^2}{m_D}(\rho_g + \frac{\lambda_3}{g_3^2}\rho_{g\lambda}). \tag{156}$$

So, for $|\rho_g + \frac{\lambda_3}{g_3^2}\rho_{g\lambda}| < 14$ the systematic uncertainty in $\eta$ is the same as the statistical error.

## 8 The theory with the $A_0$ field integrated out

In the main part of the paper we considered the 3d theory derived by dimensional reduction from the full 4d high-temperature one. In addition to the 3d Higgs field and gauge field it contains a triplet of scalar fields $A_0$. The mass of this field $\sim gT$ is larger than the typical 3d scale $\sim g^2T$, so that this field can be integrated out perturbatively. This has been done in ref.[2]. The result of this integration is a 3d gauge-Higgs system with effective parameters related to that of the original 3d theory, containing $A_0$. We refer here to the relations (I.51,I.52,I.53) of [2].

In this section, we establish the constant physics curves for a corresponding 3d lattice gauge-Higgs system, i.e. we will relate it to the original 4d high-temperature one.

The lattice action is given by (83), where all terms containing $A_0$ are omitted. The continuum action is given by (I.51). The connection between the lattice couplings $\beta_G$, $\beta_R$ and the continuum parameters $\bar{g}_3^2$ and $\bar{\lambda}_3$ is

$$\beta_G = \frac{4}{\bar{g}_3^2}\frac{1}{a}, \tag{157}$$

$$\beta_R = \frac{1}{4}\bar{\lambda}_3 a \beta_H^2 = \frac{\bar{\lambda}_3}{\bar{g}_3^2}\frac{\beta_H^2}{\beta_G}. \tag{158}$$

---

[16]There are no sizeable systematic uncertainties associated with the final volume effects, since in all cases simulations were done in such a volume that the final volume shift of $\langle R_L^2 \rangle$ was smaller than the statistical error.

In analogy with eq.(97) we write:

$$\frac{m_H^2}{4T^2} = \frac{1}{2}\gamma_0 + \frac{1}{(Ta)^2}\left[3 - \frac{1}{\beta_H} + \frac{2\beta_R}{\beta_H} - Ta\frac{f_{1m}^0}{8\pi}\Sigma\right.$$
$$\left. + (Ta)^2\frac{f_{2m}^0}{32\pi^2}\log\frac{Ta}{2} + (Ta)^2\frac{f_{2m}^0 - cP_{2m}^0}{32\pi^2}\right], \tag{159}$$

where

$$\gamma_0 = \gamma - \frac{3g^2 m_D}{16\pi T},$$
$$f_{1m}^0 = \frac{3}{2}\bar{g}_3^2 + \bar{\lambda}_3, \tag{160}$$
$$f_{2m}^0 = \frac{51}{16}\bar{g}_3^4 + 9\bar{\lambda}_3\bar{g}_3^2 - 12\bar{\lambda}_3^2,$$
$$P_{2m}^0 = \frac{51}{16}\bar{g}_3^4\eta_0 + 9\bar{\lambda}_3\bar{g}_3^2\bar{\eta} - 12\bar{\lambda}_3^2\tilde{\eta} + f_{2m}^0 c.$$

Here the parameters $\bar{\eta}$ and $\tilde{\eta}$ are the same as in the theory with the $A_0$ field, but $\eta_0$ is different from $\eta$. The value of $\eta_0$ can be defined in MC simulations in precisely the way we found $\eta$ in the previous section. We get:

$$\eta_0 = 1.62(7), \quad A = 2.62(14), \quad B = -2.05(13), \quad \tilde{\beta} = -67(20), \tag{161}$$

The fit (with $\chi^2 = 24/25$ d.o.f.) is shown on Fig. 6. We took $\mu_3 = 2.37 m_T$ for these computations [2]. The parameter $\tilde{\gamma}$ remains undetermined, as in the previous cases. Now we can, with the use of eqs. (151,152), estimate the amplitude of the 3-loop corrections to the effective potential in this theory. In complete analogy with the previous discussion we obtain

$$\tilde{\beta} = \beta - 52 \tag{162}$$

which gives $\beta = -15(20)$. The error here can certainly be decreased by the increasing statistics.

In fact, the relation between $\eta$ and $\eta_0$ can be found analytically. To this end we computed the $A_0$ contribution to the effective potential in lattice perturbation theory. The corresponding diagrams are shown in Fig. 7. The difference between the $\overline{\text{MS}}$ $A_0$ 2-loop contribution and the lattice one is found to be

$$\frac{1}{2}\frac{g_3^4}{16\pi^2}\frac{30}{16}[\log(\frac{6}{a\mu_3}) + \frac{8}{5}(\frac{\Sigma^2}{4} - \rho - \delta + \frac{5}{8}\zeta)]$$
$$+ \frac{g_3^2}{16\pi^2}6m_D^2[\log(\frac{6}{a\mu_3}) + \frac{\Sigma^2}{4} - \delta + \zeta], \tag{163}$$

where the number $\rho$ is related to a 2-loop integral defined in Appendix A and computed to be $\rho = -0.314$. This relation, together with (I.53), gives

$$\frac{81}{16}\eta = \frac{51}{16}\eta_0 + 3(\frac{\Sigma^2}{4} - \rho - \delta) + \frac{15}{8}(\log\frac{3T}{2m_D} + \frac{3}{10} + \zeta). \tag{164}$$

This establishes the required connection. With the parameter $\eta_0$ determined above we can derive an independent estimate of $\eta$:

$$\eta = 2.03(5). \tag{165}$$

This value is about 2 standard deviations below that defined in eq.(148). The difference between the two numbers may have an only statistical origin. Also, it may come from the systematic uncertainties associated with the $A_0$ field (see the discussion at the end of the previous section). In particular, if $\rho_g + \frac{\lambda_3}{g_3^2}\rho_{g\lambda} \sim 15$, the discrepancy disappears. In any case, the $2\sigma$ uncertainty in the parameter $\eta$ is 0.12, which gives already 0.3% accuracy in the determination of the temperature through the lattice parameters. In absolute units, this means $\delta T \simeq 0.5$ GeV in the vicinity of the phase transition for $m_H = 80$ GeV.

We conclude this section by noting that the 2-loop computations of the gauge invariant condensates in this theory can be extracted from sections 5 and 6 of this paper by simply dropping the $A_0$ contribution.

# 9 Conclusion

This paper is a quite technical (but absolutely necessary) step in the study of the electroweak phase transition with the use of the lattice MC methods in the framework of 3d effective theory. It provides a bridge between the lattice and continuum in terms of the constant physics curves. These curves are parametrized by 3 pure numbers ($\eta, \bar{\eta}$ and $\tilde{\eta}$ and are exact in the continuum limit. In addition, we found a number of relationships, which are exact in the continuum limit, between lattice and continuum gauge-invariant observables – condensates. The condensates were computed on the 2-loop level. This gives a possibility to study the convergence of perturbation theory in the broken phase in the vicinity of the electroweak phase transition.

The authors thank I. Montvay for many helpful discussions on dimensional reduction and on different aspects of lattice perturbation theory. K.F. is partially supported by a CEC program (CHRX - CT93 - 0319), K.R. is supported by United States Department of Energy grant DE-FG02-91ER40661.

# A  Some 1- and 2-loop computations on the lattice

In this appendix we derive the expressions for the 1-loop tadpole graph and for a number of 2-loop graphs on the lattice.

## A.1  1-loop graphs

Abbreviating the lattice propagator by

$$d(n_1, n_2, n_3, m) = \sin^2(\pi n_1/N) + \sin^2(\pi n_2/N) + \sin^2(\pi n_3/N) + (am/2)^2, \tag{166}$$

$n_i = 1, \ldots, N - 1$, we can define the lattice sum corresponding to a tadpole graph as

$$aI(N,m) = \frac{1}{4N^3} \sum_{n_i=0}^{N-1} \frac{1}{d(n_1, n_2, n_3, m)}. \tag{167}$$

In the limit $N \to \infty$ the sum can be converted into an integral,

$$\begin{aligned} aI(\infty,m) = aI(m) &= \frac{1}{4\pi^3} \int_0^\pi d^3x \frac{1}{\sin^2 x_1 + \sin^2 x_2 + \sin^2 x_3 + (am/2)^2} \\ &= \frac{1}{4} \int_0^\infty d\alpha\, e^{-\frac{1}{4}\alpha(am)^2} [e^{-\frac{1}{2}\alpha} I_0(\frac{1}{2}\alpha)]^3, \end{aligned} \tag{168}$$

where $I_0$ is the modified Bessel function. In the continuum limit $a \to 0$,

$$I(m) = \frac{1}{4\pi a}[\Sigma - am - \xi(am)^2 + 0.82\xi(am)^3 + \mathcal{O}((am)^4))] \tag{169}$$

with

$$\Sigma = \frac{8}{\pi}(18 + 12\sqrt{2} - 10\sqrt{3} - 7\sqrt{6})\mathbf{K}((2 - \sqrt{3})^2(\sqrt{3} - \sqrt{2})^2) = 3.1759114, \tag{170}$$

$\mathbf{K}$ being the complete elliptic integral of the first kind, and $\xi = 0.15281$. The values of $\xi$ and of the next coefficient are the results of a numerical computation.

A related function is the 1-loop contribution to the vacuum energy,

$$a^3 J(N,m) = \frac{1}{N^3} \sum_{n_i=0}^{N-1} \frac{1}{2} \log[d(n_1, n_2, n_3, m)]. \tag{171}$$

In the continuum limit $J(\infty, m) = J(m)$ satisfies $2dJ/dm^2 = I(m^2)$ and has the small-$a$ expansion

$$J(m) = \text{const} + \frac{1}{4\pi a}\left(\frac{1}{2}\Sigma m^2 - \frac{1}{3}am^3 - \frac{1}{4}\xi a^2 m^4 + \mathcal{O}(a^3 m^5)\right). \tag{172}$$

## A.2  2-loop graphs

The scalar sunset integral $H$ is given in eq.(I.22). Its lattice analogue is the double sum

$$\begin{aligned} H(N, m1, m2, m3) &= \frac{1}{64N^6} \sum_{n_i=0}^{N-1} \sum_{m_i=0}^{N-1} [d(n_1, n_2, n_3, m1)]^{-1} \\ &\quad [d(m_1, m_2, m_3, m2)]^{-1}[d(n_1 + m_1, n_2 + m_2, n_3 + m_3, m3)]^{-1}, \end{aligned} \tag{173}$$

where $0 \leq n_i + m_i \leq N - 1$. The $N \to \infty$ lattice limit ($N \to \infty$, $a \to 0$, $Na = $ constant) is also given by

$$H_L(m, m, m) = H(\infty, m, m, m)$$

$$
\begin{aligned}
&= \frac{1}{64\pi^6} \int_0^\pi d^3x\, d^3y\, \frac{1}{\sum \sin^2 x_i + \frac{1}{4}a^2m^2}\, \frac{1}{\sum \sin^2 y_i + \frac{1}{4}a^2m^2}\, \frac{1}{\sum \sin^2(x_i+y_i) + \frac{1}{4}a^2m^2} \\
&= \frac{1}{8} \int_0^\infty d^3\alpha\, \exp[-3(1+\frac{1}{2}a^2m^2)(\alpha_1+\alpha_2+\alpha_3)] \cdot \\
&\quad \cdot \left[ \int_0^{2\pi} \frac{dx}{2\pi} e^{\alpha_2 \cos x} I_0(\sqrt{\alpha_1^2 + \alpha_3^2 + 2\alpha_1\alpha_3 \cos x}) \right]^3 \quad (174)\\
&\approx \frac{1}{16\pi^2}\left( \log \frac{2}{am} + \frac{1}{2} + \zeta + \nu(am) + O((am)^2) \right) \\
&\equiv \frac{1}{16\pi^2}\left( \log \frac{6}{a\mu_3} + \log \frac{\mu_3}{3m} + \frac{1}{2} + \zeta + \nu(am) + O((am)^2) \right).
\end{aligned}
$$

The constants $\zeta \simeq 0.09$ and $\nu \simeq -0.6$ have been estimated by computing the integral (174) numerically and also by computing the sums for values of $N$ up to 24. Since the dependence on $am/2$ is only logarithmic, one has to go down to $am/2 \sim 0.01$ to separate the logarithmic and constant terms, and the computation is numerically demanding.

We finish this appendix by presenting two non-trivial 2-loop integrals needed for the computation of the parameter $\eta$ and for relating $\eta$ to $\eta_0$ (the parameter for the theory with the $A_0$ field integrated out). Both come from the SSV type of graphs.

The number $\delta$ is related to the derivative of the SSV diagram with respect to the scalar mass and is given by

$$\delta = \frac{1}{2\pi^4} \int_{-\frac{\pi}{2}}^{\frac{\pi}{2}} d^3x \int_{-\frac{\pi}{2}}^{\frac{\pi}{2}} d^3y \frac{\sum_i [\sin^2 x_i \sin^2(x+y)_i]}{(\sum_i \sin^2 x_i)^2 (\sum_i \sin^2(x+y)_i)(\sum_i \sin^2 y_i)} \simeq 1.94. \quad (175)$$

The number $\rho$ is related to the derivative of the SSV diagram with respect to the vector mass and is given by

$$\rho = \frac{1}{4\pi^4} \int_{-\frac{\pi}{2}}^{\frac{\pi}{2}} d^3x \int_{-\frac{\pi}{2}}^{\frac{\pi}{2}} d^3y \left\{ \frac{\sum_i [\sin^2 x_i \sin^2(x+y)_i]}{(\sum_i \sin^2 x_i)(\sum_i \sin^2(x+y)_i)} - \frac{\sum_i \sin^4 x_i}{(\sum_i \sin^2 x_i)^2} \right\} \frac{1}{(\sum_i \sin^2 y_i)^2}$$

$$\simeq -0.314. \quad (176)$$

# B   The gauge invariant effective potential

A gauge-invariant effective potential $V(\sigma)$ has been introduced in [13] by including a current $J$ coupling to $\phi^\dagger \phi$:

$$\exp[-W(J)] = \int \mathcal{D}A_i^a \mathcal{D}\phi\, \exp[-S + \int d^3x\, J\phi^\dagger\phi] \quad (177)$$

and by performing the usual transformation from $J$ to $\sigma$. Thus simply (the three-volume $V_3$ is not written explicitly)

$$W(J) = V(v(T), m_3^2 + J), \quad (178)$$

i.e. the generating functional $W(J)$ is the same as the effective potential at its minimum, but computed for $m_3^2 \to m_3^2 + J$. The relation between $\sigma$ and $J$ is given by

$$W'(J) = \frac{\partial V(v(T), m_3^2 + J)}{\partial m_3^2} \equiv \langle \phi^\dagger \phi \rangle_J = \sigma \qquad (179)$$

and the gauge-invariant effective potential is given by

$$V(\sigma) = V(v(T), m_3^2 + J(\sigma)) - J(\sigma)\sigma. \qquad (180)$$

To evaluate the Legendre transformation (180) we write $J = J_{(0)} + \hbar J_{(1)}$ ($J_{(2)}$ is not needed). To leading order in $\hbar$, it follows from eqs.(71) and (179) that

$$J_{(0)} = -m_3^2 - 2\lambda_3 \sigma \qquad (181)$$

so that from eq.(71)

$$V_{(0)}(\sigma) = V(v(T), m_3^2 + J_{(0)}) = m_3^2 \sigma + \lambda_3 \sigma^2. \qquad (182)$$

To order $\hbar$

$$J_{(1)} = \frac{\lambda_3}{2\pi}\left[\bar{m}_1 + \frac{3g_3^2}{8\lambda_3}(2\bar{m}_T + \bar{m}_L)\right], \qquad (183)$$

where now

$$\bar{m}_1 = \sqrt{4\lambda_3 \sigma}, \quad \bar{m}_T = \sqrt{\frac{1}{2}g_3^2 \sigma}, \quad \bar{m}_L = \sqrt{m_D^2 + \frac{1}{2}g_3^2 \sigma}. \qquad (184)$$

These equations show that the calculation is valid for $\sigma > 0$ or for $m_3^2 + J < 0$, i.e. in the broken phase. With these masses, the 2-loop result for the gauge invariant effective potential is

$$V(\sigma > 0) = m_3^2 \sigma + \lambda_3 \sigma^2 - \frac{\hbar}{12\pi}(6\bar{m}_T^3 + 3\bar{m}_L^3 + \bar{m}_1^3) \qquad (185)$$
$$- \frac{\hbar^2 \lambda_3}{64\pi^2}\bar{m}_1[5\bar{m}_1 + \frac{3g_3^2}{2\lambda_3}(2\bar{m}_T + \bar{m}_L)] + \hbar^2 V_2(\bar{m}_T, \bar{m}_L, \bar{m}_1, m_2 = 0),$$

where $V_2$ is given in eq.(34).

If $\sigma < 0$ or $m_3^2 + J > 0$ one has to use the eqs. (73), (74) and (40), which are appropriate for the symmetric saddle point. Now the leading term in $V(v)$ in eq.(40) is $\mathcal{O}(\hbar)$ and one obtains

$$\sigma(J) = -\frac{\hbar}{2\pi}\sqrt{m_3^2 + J}. \qquad (186)$$

The same procedure as before leads to

$$V(\sigma < 0) = \left(m_3^2 - \frac{3\hbar}{8\pi}g_3^2 m_D\right)\sigma + \frac{3}{4}\left(g_3^2 \log\frac{\mu_3 \hbar}{-4\pi\sigma} + \frac{1}{4}g_3^2 + 2\lambda_3\right)\sigma^2 - \frac{4\pi^2}{3\hbar^2}\sigma^3, \qquad (187)$$

neglecting a $\sigma$-independent part depending on $m_D$, $\log(\mu_3/m_D)$, and the coupling constants. Note the dependence on $\hbar$.

Although explicit forms of the gauge-invariant potential have been given here one has not gained anything of practical use. One is anyway only interested in the physical ground-state value of $\langle \phi^\dagger \phi \rangle$ and this is already given by eqs.(76-78).

# C  Distribution of $\langle \phi^\dagger \phi \rangle$ in a finite system

In the course of lattice simulations $R^2 = \frac{1}{V_3}\int d^3x \phi^\dagger(x)\phi(x)$ will be studied in finite volume systems and it is of great use to know its distribution $dN/dR^2$ in the continuum, but in a finite 3-volume $V_3$. By definition we have, in terms of the functional integral in eq.(38),

$$\begin{aligned}
\frac{dN}{dR^2} &= \int \mathcal{D}\psi\, \delta(R^2 - \frac{1}{V_3}\int d^3x \phi^\dagger(x)\phi(x))e^{-S[\psi]} \\
&= \int_{-\infty}^{\infty} \frac{dt}{2\pi} \int \mathcal{D}\psi\, \exp(-S[\psi] + itR^2 - it\frac{1}{V_3}\int d^3x \phi^\dagger \phi) \\
&= \int_{-\infty}^{\infty} \frac{dt}{2\pi} \exp\{itR^2 - V_3 V[v(T), m_3^2 + it/V_3]\}.
\end{aligned} \qquad (188)$$

The integral over $t$ is computed with the saddle point method: the saddle point $t_s$ is determined by

$$R^2 = \frac{dV[v(T), m_3^2 + it_s/V_3]}{dm_3^2} = \frac{dV[v(T), m_3^2]}{dm_3^2} + i\frac{t_s}{V_3}\frac{d^2V[v(T), m_3^2]}{(dm_3^2)^2} + \ldots, \qquad (189)$$

so that the saddle point is at

$$\frac{it_s}{V_3} = \frac{R^2 - V'[m_3^2]}{V''[m_3^2]}, \qquad (190)$$

where the derivatives always are with respect to $m_3^2$, keeping $\phi = v(T)$. Near the average

$$\langle R^2 \rangle = V'(m_3^2), \qquad (191)$$

$t_s/V_3$ is small relative to $m_3^2$ and it is consistent to expand as in eq.(189). For the distribution one obtains

$$\frac{dN}{dR^2} = \frac{e^{-V_3 V[m_3^2]}}{\sqrt{-2\pi V''(m_3^2)/V_3}} \exp\left\{\frac{V_3}{2V''(m_3^2)}[R^2 - V'(m_3^2)]^2\right\}. \qquad (192)$$

This is a Gaussian with the average as in eq.(191) and the width $-V''(m_3^2)/V_3$. Note that $V''(m_3^2) < 0$: for a simple tree potential near a broken minimum ($m_3^2 < 0$) the derivatives are $V(m_3^2) = -m_3^4/(4\lambda_3)$, $V'(m_3^2) = -m_3^2/(2\lambda_3)$, $V''(m_3^2) = -1/(2\lambda_3)$. Eq.(192) shows how the fluctuations freeze to zero in the infinite volume limit.

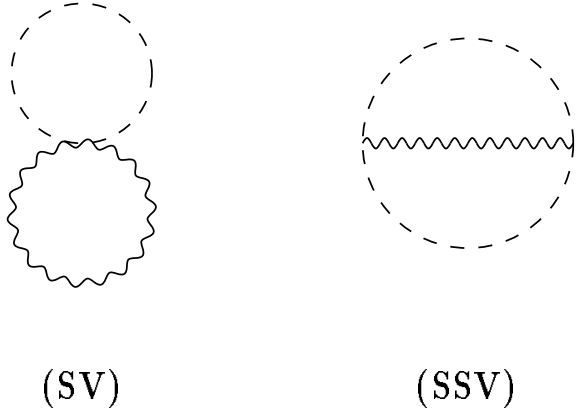

Figure 1: The vacuum 2-loop diagrams used in the computation of $\bar{\eta}$.

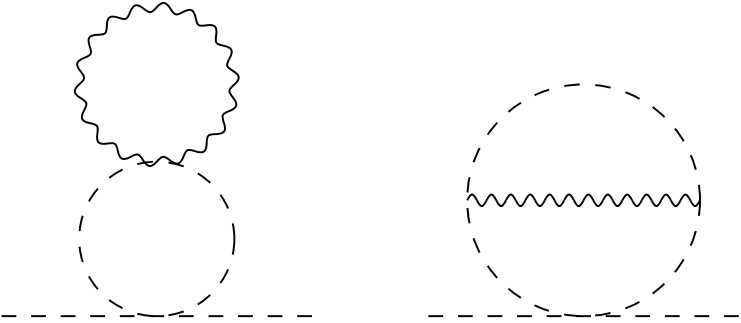

Figure 2: The Higgs self-energy 2-loop diagrams used in the computation of $\bar{\eta}$.

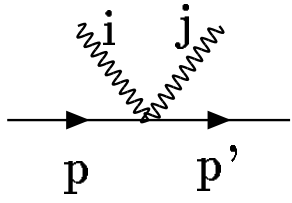    $\delta_{ij} \cos \frac{1}{2}a(p_i + p'_i)$

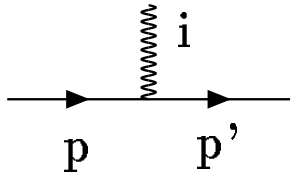    $\frac{2}{a} \sin \frac{1}{2}a(p_i + p'_i)$

Figure 3: The scalar-gauge vertices in lattice perturbation theory.

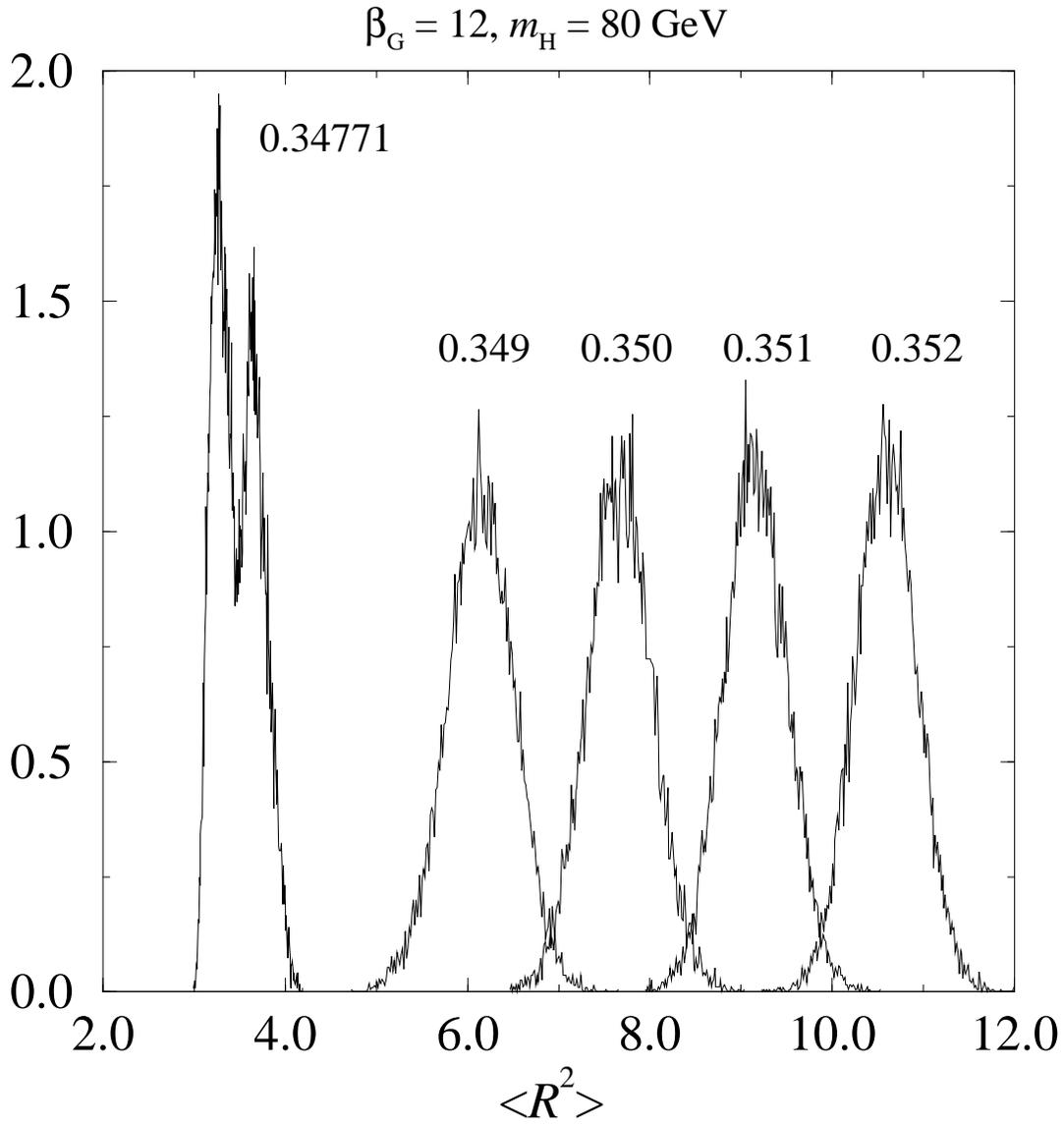

Figure 4: The evolution of the distribution of $\langle R_L^2 \rangle$ with $\beta_H$.

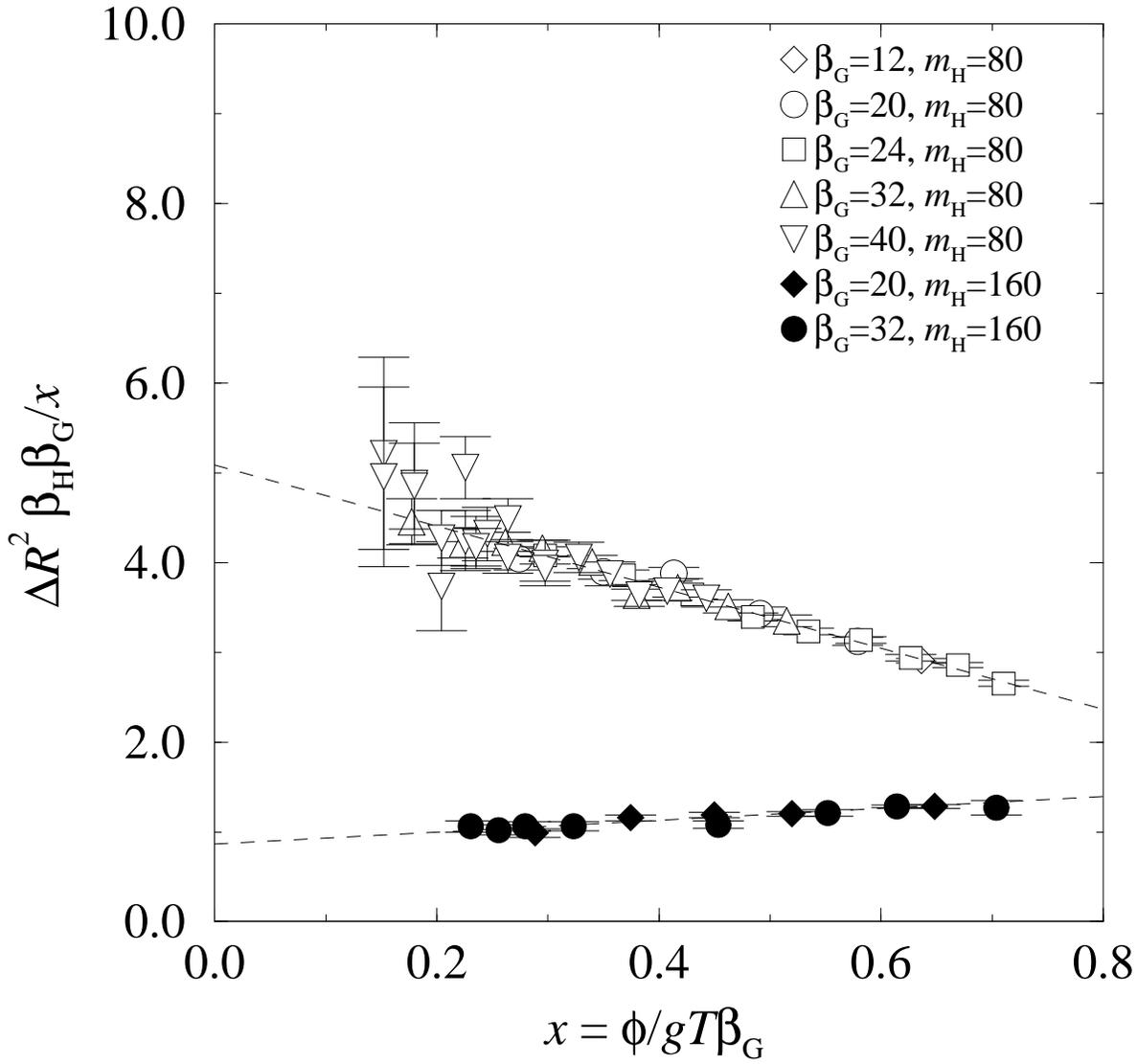

Figure 5: The quantity $\frac{2\Delta}{am_T}$ as a function of $x = am_T/2$.

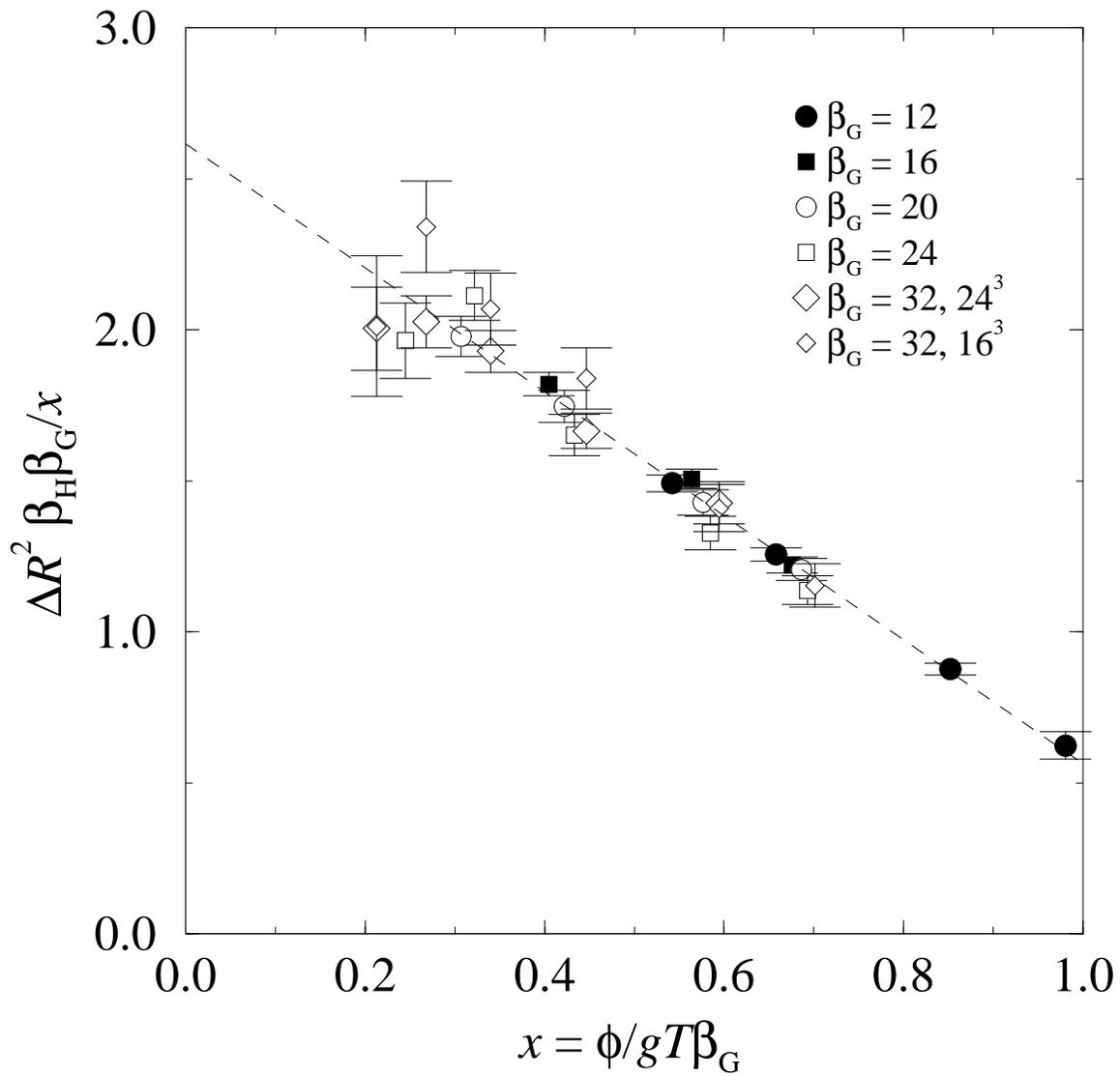

Figure 6: The same as Fig.5, for a theory without $A_0$ field.

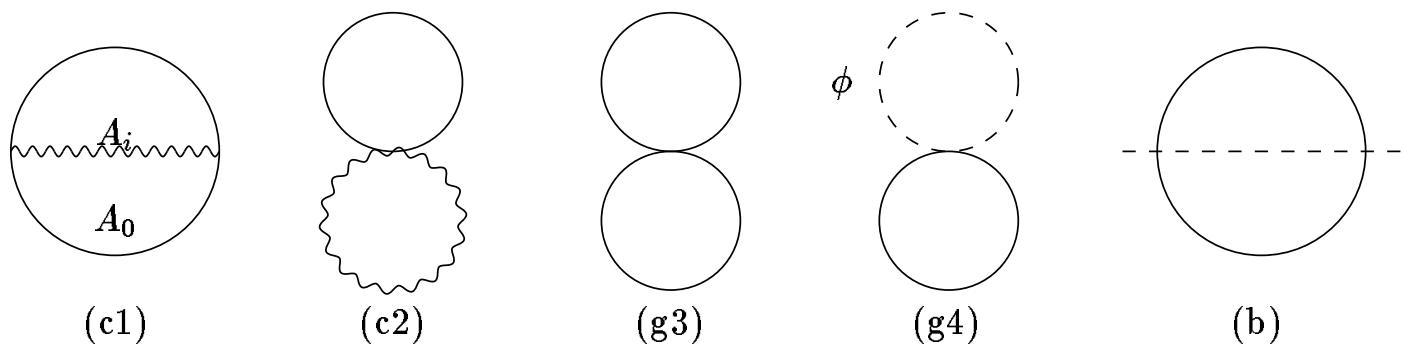

Figure 7: The $A_0$ contribution to the effective potential.